\documentclass[lettersize,journal]{IEEEtran}
\usepackage{bm}
\usepackage{cite} 
\usepackage{amssymb}
\usepackage{graphicx}
\usepackage{subfigure}
\usepackage{algorithmic}
\usepackage[english]{babel}
\usepackage[utf8]{inputenc}
\usepackage{graphicx}
\usepackage{graphics}
\usepackage{textcomp}
\usepackage{subfigure}
\usepackage{tabularx}
\usepackage{times}
\usepackage{multirow}
\usepackage{makecell}
\usepackage{bm}
\usepackage{latexsym}
\usepackage{booktabs}
\usepackage{threeparttable}
\usepackage{epsf}
\usepackage[ruled]{algorithm2e}
\usepackage{color,colortbl}
\usepackage{float}
\usepackage[table,xcdraw]{xcolor}
\usepackage{booktabs}
\usepackage{extarrows}
\usepackage{setspace}
\IEEEoverridecommandlockouts

\begin{document}
	% towards WNCS, goal-oriented semantic analysis and communication-control codesign
	\title{Goal-Oriented Communication, Estimation, and Control over Bidirectional Wireless Links}
	
	\author{Jie Cao,~\IEEEmembership{Member,~IEEE},
		Ernest Kurniawan,~\IEEEmembership{Senior Member,~IEEE},	
		Amnart Boonkajay,~\IEEEmembership{Member,~IEEE},\\
		Nikolaos Pappas,~\IEEEmembership{Senior Member,~IEEE},	
		Sumei Sun,~\IEEEmembership{Fellow,~IEEE},
		Petar Popovski,~\IEEEmembership{Fellow,~IEEE}
		%		Institute for Infocomm Research (I$^2$R), A*STAR, Singapore\\
%		\thanks{This work is a part of project “Future Proof Reliable and Resilient Wireless Communications for Virtual Power Plant (W-VPP),” supported by National Research Foundation (NRF) Singapore, under its Industry Alignment Fund (Pre-Positioning) (IAF-PP) for Urban Solutions and Sustainability (USS) Domain, Research Innovation Enterprise 2020 Plan (RIE2020), and Energy Grid 2.0 Programme. 	{Email: %cao\_jie@i2r.a-star.edu.sg, 
%				ekurniawan@i2r.a-star.edu.sg
%				%	,	amnart\_boonkajay@i2r.a-star.edu.sg, nikolaos.pappas@liu.se, sunsm@i2r.a-star.edu.sg and petarp@es.aau.dk.
%		}}
%		\thanks{ Part of this work has been accepted for 2023 IEEE Globecom\cite{globecom23_CJ}.}
				\thanks{Jie Cao is with the School of Electronic and Information Engineering, Harbin Institute of Technology, Shenzhen
			518055, China (Corresponding author: Jie Cao, e-mail: caojhitsz@ieee.org).}\\
		\thanks{ Ernest Kurniawan,  Boonkajay Amnart and Sumei Sun are with the Institute of Infocomm Research, Agency for Science, Technology and Research, Singapore 138632.}
				\thanks{Nikolaos Pappas is with the Department of Computer and Information Science, Linköping University, 58183 Linköping, Sweden.}
		\thanks{Petar Popovski is with the Department of Electronic Systems, Aalborg	University, Danish.}
		\thanks{ Part of this work has been accepted for 2023 IEEE Globecom\cite{globecom23_CJ}.}
	}

	\maketitle
	
	\begin{abstract}
		We consider a wireless networked control system (WNCS) with bidirectional imperfect links for real-time applications such as smart grids. To maintain the stability of WNCS, {captured by the probability that plant state violates preset values}, at minimal cost, heterogeneous physical processes are monitored by multiple sensors. This status information, such as dynamic plant state and Markov Process-based context information, is then received/estimated by the controller for remote control. However, scheduling multiple sensors and designing the controller with limited resources is challenging due to their coupling, delay, and transmission loss. We formulate a Constrained Markov Decision Problem (CMDP) to minimize violation probability with cost constraints. We reveal the relationship between the goal and different updating actions by analyzing the significance of information that incorporates goal-related usefulness and contextual importance. Subsequently, a goal-oriented deterministic scheduling policy is proposed. Two sensing-assisted control strategies and a control-aware estimation policy are proposed to improve the violation probability-cost tradeoff, integrated with the scheduling policy to form a goal-oriented co-design framework. Additionally, we explore retransmission in downlink transmission and qualitatively analyze its preference scenario. Simulation results demonstrate that the proposed goal-oriented co-design policy outperforms previous work in simultaneously reducing violation probability and cost.
		
	\end{abstract}
	
	
	\section{Introduction}
	Wireless networked control systems (WNCSs) are receiving increasing attention and are expected to propel the development of emerging cyber-physical and real-time applications such as smart grid, factory automation, and {autonomous robot}\cite{globecom23_CJ,8166737}. As shown in Fig. 1, a typical WNCS comprises a plant with dynamic states, multiple observation sensors, a remote control controller, and a set of actuators. Due to the coupling of these elements, the design of sensing, communication, and control is, therefore, a challenge. To accomplish specific tasks (\emph{e.g.}, accurate tracking or stable control) in WNCSs\cite{10012674, secon2023}, information distilled from observations and control commands is required to be fresh, accurate,  and useful to the corresponding goals.
	
	{However, heterogeneous sensors observing different physical processes contain information with distinct significance, affecting the goal in different manners\cite{2023arXiv230304908F}. For example, two sensors in Fig. 1 capture the state of a dynamic physical process (\emph{e.g.}, plant state) and a random process (\emph{e.g.}, environment change), respectively. These heterogeneous observations need to be transmitted in a timely manner for calculating control commands to ensure system stability. However, each observation has a different and non-trivial impact on this goal. Therefore, it is challenging to schedule multiple packets efficiently over a wireless channel with limited resources\cite{8558500}. The controller is also expected to efficiently transmit control commands to the actuator based on the received packets, thereby accomplishing the goal. These motivate the exploration of the non-trivial relationship between the goal and heterogeneous observations/control commands and the development of the goal-oriented co-design in WNCSs.}
	
	\begin{figure}[tbp]
		\centering
		{\includegraphics[height=3.6cm]{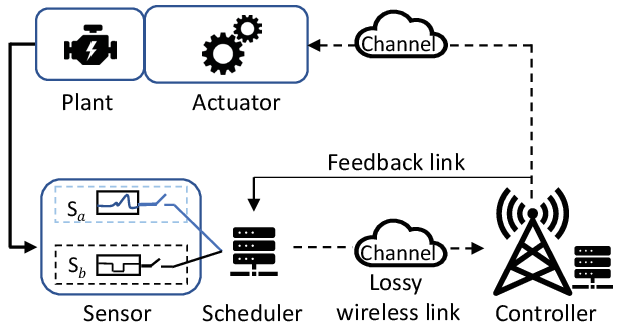}}
		\caption{A closed-loop WNCS with heterogeneous physical processes.}\label{fig_1}
	\end{figure}
	
	{To evaluate the importance and significance of transmitted packets, various goal-oriented performance metrics such as the age of information (AoI)\cite{6195689} and its variants\cite{8845114,9024463,9137714}  have been proposed. Based on these metrics, several sampling and scheduling policies have been explored to improve transmission efficiency for the communication resource-limited single-source system   \cite{8845114,9024463,9137714,8995639,9162973,9836031,2022arXiv220708996S,9695972,2023arXiv230300507N,9146773,10181305,9551200, ICCW23, JCN23}.
		Furthermore, communication and control co-design has been investigated without considering the goal of WNCS and its interplay between communication and control actions\cite{9478879,8629300,9369024,10004994,9690022,9052443,9614347, 8734802,9729746}. Moreover, the coupling of sensing, communication, and control has yet to be explored in depth in the WNCS\cite{10034831,9493202,9599512,9807392,9305697,8865111,10202236}. Therefore, further investigation is required for goal-oriented co-design in Wireless Networked Control Systems (WNCSs).}
	
	%		\subsection{Contribution}
	Motivated by the above open issues, we consider a WNCS with multiple sensors to monitor heterogeneous physical processes and a remote controller to maintain the plant state over bidirectional imperfect links.
	To ensure the stability of WNCSs at minimal cost, we aim to reduce the number of control overshoots, with the consideration of heterogeneous action costs.
	% the extreme case dominates the system downtime.
	Hence, the violation probability of the plant state against the preset threshold is adopted as the primary KPI. 
	Then, the scheduler and controller are designed to address the following questions:
	\textit{What is the relationship between multiple observations/control packets with different significance and the specific goal? How do we jointly design the sensing, scheduling, and control strategies with limited resources to improve the tradeoff between violation probability and cost?}

	The main contributions are as follows.
	\begin{itemize}
		\item We first reveal the relationship between the WNCS goal and the actions of the scheduler/controller in a WNCS with bidirectional imperfect links. Goal-related dynamic state and Markov process-based context information are monitored simultaneously. Furthermore, the event-triggered controller is adopted to reduce the control cost.
		For the  goal of WNCSs, we investigate the 	
		significance of information that incorporates goal-oriented usefulness and contextual importance.
		However, the existing works either ignore the significance of the information concerning the goal and context\cite{6195689,8845114,9137714,9146773} or focus on the case of single source with one-way communication only\cite{globecom23_CJ,9146773,2023arXiv230304908F}.
		\item 
		{CMDP-based scheduling and control problems are} formulated to ensure the stability of WNCS with a given cost, in which state violation probability is minimized, and heterogeneous action costs are considered.  
		The scheduler decides which sensor is scheduled or remains idle at each time slot based on the goal. 
		The event-triggered controller decides whether to control based on the received plant state and context information.
		Nevertheless, only the scheduling scheme is proposed in the existing AoI-oriented system \cite{8995639,9162973}. In addition, the consideration of goal in the previous co-design work is limited \cite{8865111}.

		\item 
		A goal-oriented estimation, scheduling, and control co-design policy is proposed to improve the violation probability-cost tradeoff. 
		The formulated CMDP is transformed into an MDP based on the analyzed structural results using the Lagrangian relaxation method. Then, a deterministic scheduling policy is proposed using the relative value iteration algorithm (RVIA) and bisection method. Furthermore, control-aware estimation, sensing-assisted control, and state-dependent retransmission policies are proposed for performance improvement.
		
		\item 		 
		The simulation results demonstrate that the proposed goal-oriented co-design policy can significantly reduce the violation probability and total cost compared to the existing work\cite{8995639,9162973,8845114,9137714}.
		Moreover, the analysis of the impact of system parameters on the proposed algorithms in different scenarios is verified by simulation results. 
	\end{itemize}

	%	\textit{Organization:}
	%	%The rest of this paper is organized as follows.
	%	Sections II and III  introduce the considered WNCS model and formulated problem. CMDP-based scheduling policy is presented in Section IV. Goal-oriented  communication and control co-design  is provided in Section V.  Conclusion and future work are presented in Section VI.
	
	\textit{Notation:}
	Scalar quantities are represented using lowercase letters. Vectors and matrices are denoted by bold lowercase and uppercase fonts, respectively. Sets are denoted using italic uppercase letters. Table I lists the notations.

	\renewcommand{\arraystretch}{1.2}
	\begin{table}[ht]
		\centering
		\caption{Definitions of the key variables in the paper}
		\footnotesize
		\begin{tabular}{lllll}
			\cline{1-2}
			\textbf{Notations}                                & 	\textbf{Definition(s)}  &  &  &  \\ 	\cline{1-2}
			$\mathbf{A}$,  $\mathbf{B}$, $\mathbf{C}$             & System, input, and measurement matrices  &  &  &  \\ 	\cline{1-2}
			$\mathbf{x}$, $\mathbf{u}$, $y$  & Plant state, control input, and measurement     &  &  &  \\ \cline{1-2}
			$m$, $n$                         & Dimension of system and input matrices    &  &  &  \\ 	\cline{1-2}
			$k$, $K$  & Time index and total number of slots   &  &  &  \\ \cline{1-2}
			$\hat{\mathbf{x}}$, ${\mathbf{e}}$  & Estimation and estimation error  of plant state    &  &  &  \\ 	\cline{1-2}
			$\mathbf{\Theta}$, $\mathbf{\Phi}$ & Covariance of estimation error and plant state     &  &  &  \\ 	\cline{1-2}
			$\mathbf{w}$, $\mathbf{R}_w$                      & Noise and noise variance    &  &  &  \\ 	\cline{1-2}
			$p$, $\bar{p}$                   & Transition  and self-transition  probabilities of Markov chain    &  &  &  \\ 	\cline{1-2}
			$P_{i,j}$                      &  Transition probability from state $i$ to $j$ of Markov chain   &  &  &  \\ 	\cline{1-2}
			$v$, $\hat{v}$             & DTMC based environment condition  and its estimation  &  &  &  \\ 	\cline{1-2}
			%	\textcolor{red}{$o$, $a$, $b$, $c$}              & Subscript of device, sensor $a$, sensor $b$, controller  &  &  &  \\ 	\cline{1-2}
			$\Upsilon$             & Estimation quality of context information   &  &  &  \\ 	\cline{1-2}
			$\epsilon$             & Packet error probability of wireless transmission    &  &  &  \\ 	\cline{1-2}	
			$\delta$, 	$\xi$               & Indicator of decision/successful transmission&  &  &  \\ 	\cline{1-2}	
			$\Delta$               & AoI of the received sensor's packet at the controller&  &  &  \\ 	\cline{1-2}	     
			$\zeta$, $J$, {$c$}               & Preset threshold, indicator of violation event, {action cost}&  &  &  \\ 	\cline{1-2}	
			$s$, $\mathcal{S}$, {$\alpha$}, $\mathcal{A}$                 & State, state space, action, {action} space of CMDP&  &  &  \\ 	\cline{1-2}	 
			$\mathcal{T}$, $\mathcal{R}$               & State transition matrix and reward of CMDP&  &  &  \\ 	\cline{1-2}	  
		\end{tabular}
	\end{table}
	
	\section{Related Work}
	This section reviews the related work on goal-oriented communication and co-design in WNCSs.
	
	\subsection{Goal-Oriented Communication}

	To support real-time applications in WNCSs, wireless network key performance indicators (KPIs) such as throughput, delay, and packet drop rate have been optimized to enable ultra-reliable and low-latency communication\cite{9690057}. However, the conventional KPIs ignore the semantic attributes of transmitted information, which treat all the packets equally and lead to less efficiency in achieving the specific goal\cite{9955525}. To address this, AoI has been proposed to capture the information freshness, which is one of the critical semantic attributes\cite{6195689}. Since AoI can mathematically characterize the impacts of packet loss and delay on the state estimation error\cite{8845114}, many AoI-based sampling and scheduling policies have been proposed to facilitate remote estimation in WNCSs. In \cite{8995639} and \cite{9162973}, time slot-based and random access-based scheduling algorithms were proposed to minimize the average AoI, respectively. The sampling rate, computing scheduling, and transmit power were jointly optimized for the AoI minimization in WNCSs\cite{9836031}. In \cite{2022arXiv220708996S}, a transmission scheduling policy was proposed to minimize the average number of transmissions subject to an average AoI constraint. The update interval was also optimized for minimizing AoI in random access-based Internet of Things (IoT) systems \cite{9695972}.

	However, the aforementioned AoI-based work fails to evaluate packets based on the content and significance of information associated with the goal, even though the AoI can capture the timeliness of  packets\cite{9551200}.  
	Therefore, various AoI variants, such as value of information\cite{8845114}, age of incorrect information (AoII)\cite{9024463,9137714},  urgency of information\cite{9146773}, and {age of actuation}\cite{2023arXiv230300507N} have been proposed to quantify the significance of information based on the communication goal, context and semantic mismatch.
	In  \cite{10181305}, the weighted AoII and throughput were optimized jointly to support the coexistence of task-oriented and data-oriented communications.
	However, exploring the relationship between goal and communication policy remains unclear.  
	Based on the significance and effectiveness of information, 	 a goal-oriented sampling and communication policy was proposed to reduce real-time reconstruction error and the cost of actuation errors\cite{9551200}.

	{Overall, goal-oriented communication, the basis for pragmatic communication,  has attracted a lot of attention\cite{petar_arxiv}. Many semantic attributes have been introduced and adopted to design sampling and scheduling policies for a goal accomplishment\cite{8845114,9024463,9137714,8995639,9162973,9836031,2022arXiv220708996S,9695972,9146773,2023arXiv230300507N,10181305,9551200}.  
		However, the goal-oriented WNCS has not been sufficiently investigated, especially for the closed-loop WNCS with heterogeneous traffic. } 
	Scheduling multiple packets with different significance and contextual importance to the specific goal remains unexplored.

	\subsection{Communication and Control Co-Design}
	
	On the control side of WNCS, control law is designed to achieve specific control goals over wireless networks. For example, the steady-state error is minimized for automatic tracking, and the overshoot is minimized to avoid failure and system downtime\cite{9478879}. 
	To achieve specific control goals,  the classical controller transmits control commands periodically. The event-triggered controller generates control actuation reactively when the plant state violates a certain threshold\cite{8629300}. 
	In addition to the different types of controllers mentioned above,	several control policies have been proposed to achieve the desired stability and control performance, such as proportional-integral-derivative  control\cite{8166737}, linear quadratic regulator control\cite{9369024}, and {model predictive control (MPC)}\cite{10004994}.
	However, network delay and loss in WNCSs may degrade the control performance and destabilize the system, which were not considered in the work above.

	More recently, communication and control co-design has garnered substantial attention, as it can eliminate the effects of limited throughput and poor link quality of wireless networks.
	In \cite{9690022},   a  dynamic transmission scheduling strategy was proposed to optimize control performance by allocating network resources based on link quality predictions.
	Motivated by the tradeoff between control performance and energy cost,  a state-dependent scheduling algorithm was proposed to find the optimal power allocation policy\cite{9052443}.
	In \cite{9614347}, a greedy state-error-dependent scheduling policy was proposed to achieve a minimum average linear quadratic cost.
	For saving energy, a novel co-design approach for distributed self-triggered control over wireless multi-hop networks was proposed\cite{8734802}. In \cite{9729746}, an integrated scheduling
	method of sensing, communication, and control for backhaul
	transmission was proposed for unmanned aerial vehicle (UAV)  networks.
	
	The aforementioned work on communication and control co-design focused on either the control design over communication\cite{8629300,9369024,10004994} or network resources allocation tailored for control performance\cite{9690022,9052443,9614347,   8734802,9729746}. 
	However, the goal of WNCS and its interplay between communication and control actions were not analyzed. Moreover, the semantic attributes of transmitted information were not considered\cite{8629300,9369024,10004994,9690022,9052443,9614347,   8734802,9729746}.

	\subsection{Goal-Oriented Co-Design} 
	
	Due to the limited resources and unclear interplay among multiple subsystems in a WNCS, it is still challenging to schedule sensing and control packets efficiently based on the significance of goal-related information\cite{8558500}. Our previous work investigated the communication and control co-design for the WNCS with a single imperfect link and homogeneous action cost\cite{globecom23_CJ}. We have also introduced the age of loop to quantify the timeliness from the sensor to the actuator\cite{10012674}. Similarly,  a full loop AoI-based control and transmission co-design architecture was proposed for industrial cyber-physical systems\cite{10034831}. In \cite{9493202}, a WNCS with two-way imperfect communications was considered, and an AoI-aware co-design of scheduling and control was investigated. In \cite{9599512}, a robust distributed MPC scheme based on AoI was established. In \cite{9807392}, AoI was adopted for the event-triggered communication scheduler to manage the information updating process. In \cite{9305697}, the data interarrival rate and code length were jointly designed for AoI-oriented IoT systems. Considering the goal of WNCS,	 an uplink–downlink transmission scheduling problem was addressed to minimize the mean square of error (MSE) of plant state in  \cite{8865111}. 
	In \cite{10202236}, an AoI-aware communication and control co-design scheme was proposed to improve the control performance.
	
	The existing AoI-based and MSE-based scheduling policies have illuminated the path for goal-oriented design, but only a single attribute of the information was considered, which may lead to inefficient transmission.
	In \cite{9165797}     and \cite{10138567},  goal/task-oriented task offloading policies were proposed to attain a better tradeoff between computation latency and energy consumption.
	In \cite{10005612}, a quality-of-service-oriented sensing-communication-control co-design scheme was proposed for UAV  positioning systems.
	In \cite{10049657}, the features of tasks were considered, based on which   a task-oriented
	prediction and communication co-design framework was proposed.

	Although recent advances in wireless sensing and control have facilitated the co-design of multiple subsystems, the coupling of sensing, communication, and control has not been explored in depth in the WNCS. Moreover, the interplay between the goal and observations/control commands has not been revealed clearly. Therefore, it is worth conducting further research on the goal-oriented co-design in WNCSs.

	\section{System Model}
	In this section, we introduce the considered WNCS, including the heterogeneous physical process model, wireless transmission, and control models.
	
	\subsection{System Description}
	We consider a WNCS with two-way communications and heterogeneous physical processes to be monitored,  as shown in Fig. \ref{fig_1}. 
	The considered WNCS consists of two different sensors, a scheduler, a controller, and an actuator. 
	This can be employed in emerging mission-critical applications such as load frequency control (LFC) in smart grid\cite{9807392}.

	The specific information flow and assumptions of the considered WNCS are presented below.
	\begin{enumerate}
		\item  We consider a time-slotted WNCS, where sensors obtain observations, and the controller generates control commands at the beginning of each time slot\cite{8865111}. 
		\item  The dynamic physical  process (\emph{e.g.}, plant state) and random process (\emph{e.g.}, environment change) are monitored by sensors $S_a$ and $S_b$, respectively. 
		\item  The scheduler selects one sensor and transmits its update to the controller through the wireless uplink channel. Non-orthogonal transmission is not considered in this paper. Due to the limited resources, we assume that only one sensor can be scheduled in each time slot\cite{8865111,2023arXiv230304908F}. 
		\item  The controller is assumed to be event-triggered and sends 
		the control command to the actuator\cite{8845114} when the plant state violates the preset threshold\cite{9807392}.
		\item  {For the uplink (scheduler-controller link), we assume that the sensing packet is transmitted with one slot delay and probability of transmission error, without retransmission\cite{2023arXiv230304908F}. For the downlink (controller-actuator link), the control command with limited data size is assumed to be transmitted instantly with a transmission error probability, following \cite{8865111}. }
		\item The receiver sends an acknowledgment
		(ACK)/negative ACK (NACK) for successful/failed transmissions instantly without error.
		
		%Besides, the sensors and controller consume a cost $c$ for each transmission action.
	\end{enumerate}

	\subsection{Dynamic Process Model}
	{We consider heterogeneous traffic in the WNCS, including  dynamic physical process monitored by  sensor $S_a$ and random Markov process monitored by sensor $S_b$. }  
	These two sources affect  the goal of WNCS in different ways, leading to their different importance and significance.
	\subsubsection{Linear-Time-Invariant (LTI System}
	We consider the dynamic physical process (\emph{e.g.}, frequency deviation in LFC system) modeled by the discretized LTI system\cite{8845114} 
	\begin{equation}
		\begin{aligned}
			&\mathbf{x}_{k+1}=\mathbf{A}\mathbf{x}_k+\mathbf{B}\mathbf{u}_{k}+\mathbf{w}_k,\\
			&{y}_k=\mathbf{C}\mathbf{x}_k,\label{eq_1}
		\end{aligned}
	\end{equation}
	where the subscript $k$ represents the time index for plant state evolution. 
	$\mathbf{A}\in\mathbb{R}^{m\times m}$ and $\mathbf{B}\in\mathbb{R}^{m\times n}$ represent the system matrix and input matrix, respectively,
	$\mathbf{C}\in\mathbb{R}^{1\times m}$ is the measurement matrix.
	Besides, $\mathbf{x}_k\in\mathbb{R}^{m\times1}$ is the plant state, 	$\mathbf{u}_k\in\mathbb{R}^{n\times1}$ denotes the control input, and  ${y}_k $ is the measurement. $\mathbf{w}_k\in\mathbb{R}^{m\times 1}$ is the normally distributed noise with the distribution $\mathcal{N}(\textbf{0},\mathbf{R_w})$, where $\mathbf{R_w}$ is the noise variance.
	To facilitate presentation, we ignore the measurement equation  and focus on the state equation in the analysis part.
	The state is assumed to be fully measured.
	Generally, $\mathbf{x}$  represents the difference  between the true value and the preset value of the considered process. Hence, the goal of the system is to
	maintain the state $\mathbf{x}$  close to $\textbf{0}$.
	{The sensor used to monitor the plant state is denoted as sensor $S_a$.}

	\subsubsection{Discrete-Time Markov Chain (DTMC)} 
	The physical dynamics in real systems may suffer from abrupt variations (\emph{e.g.}, environment changes).
	The state of the environment provides important context information, which may affect the sensitivity and tolerance of the equipment to the plant state.
	This random process can be modeled as a DTMC and is assumed to be ergodic\cite{9382948}. 
	
	\begin{figure}[htbp]
		\centering
		{\includegraphics[height=2cm]{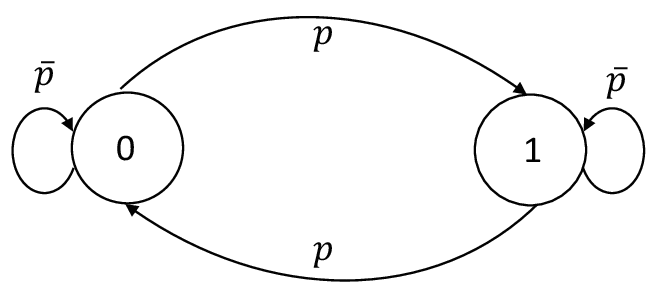}}
		\caption{Markov chain of environment/condition changes.}\label{fig_2}
	\end{figure}

	The state of the environment is denoted by $v$ and takes value from 0 and 1, as shown in Fig. \ref{fig_2}.  The transition matrix is described by $P_{i,j}=\Pr\left\{v_{k+1}=j|v_k=i\right\}$, where $v_k$ is the environment condition at time slot $k$.
	We have $P_{0,0}=P_{1,1}=\bar{p}$ and $P_{1,0}=P_{0,1}=p=1-\bar{p}$, where $p$ and $\bar{p}$ denote the self-transition and transition probabilities, respectively.   {The sensor used to monitor context information is denoted as sensor $S_b$. }

	\subsection{Wireless Transmission and Control Model}
	We consider the wireless erasure channel in  WNCSs,  and assume that channel state remains unchanged within one time slot and varies independently among different slots.
	The packet error	probability of wireless transmission is denoted as $\epsilon\in(0,1)$. 
	The sensors generate status updates $\mathbf{x}_k$ and $v_k$ by sampling
	the source at each time slot, where the sampling process is assumed to be without incurring  any latency and cost\cite{8865111}. 
	To alleviate the effect of limited communication bandwidth,	the controller is assumed to be event-triggered and generates control command $\mathbf{u}_k$ at the triggered time slot\cite{9807392}. 
	Also, we assume that the system is controllable but unstable without control, \emph{i.e.,} $\rho(\mathbf{A})>1$, where $\rho(\mathbf{A})$ is the spectral radius of system matrix $\mathbf{A}$\cite{8865111,9807392}.
	
	The decision to sample and transmit for the sensor and controller    at time
	slot $k$ is denoted by $\delta_{o,k}\in\left\{1,0\right\}$, where $o\in\left\{a,b,c\right\}$ represents  sensors $S_a$,  $S_b$, and controller, respectively. Also, let $\xi_{o,k}\in\left\{1,0\right\}$ represent  the outcome of the transmission.
	Let $\xi_{o,k}=1$ indicate a successful transmission with probability ${\Pr}\left[\xi_{o,k}=1|\delta_{o,k}=1\right]=1-\epsilon_{o}$.

	\section{Problem Formulation and Analysis}
	This section  presents the evolution of the plant state  with respect to time slots and  actions of the scheduler and controller.
	Then the goal of the considered system, \emph{i.e.}, violation probability, is analyzed, based on which a joint optimization problem is formulated.

	\subsection{State Estimation by the Controller }

	Recall that  each uplink transmission process takes a time slot to be performed.
	Hence, the controller receives an update from the sensor
	with one slot delay.
	To generate effective control commands, the controller must
	maintain accurate estimates of plant state and context information (e.g., environment/condition changes) at each time slot. 
	\subsubsection{LTI System}	
	Sensor $S_a$ monitors  the LTI-based plant state to obtain the goal-related value at time slot $k$. 
	The plant state-estimator at the controller is given as 
	\begin{equation}
		\hat{\mathbf{x}}_{k+1} = \begin{cases}
			\mathbf{A}{\mathbf{x}}_{k}+\mathbf{B}\mathbf{u}_{k}, & \delta_{a,k}\xi_{a,k}=1, \\
			\mathbf{A}\hat{\mathbf{x}}_{k}+\mathbf{B}\mathbf{u}_{k}, & otherwise.
		\end{cases}\label{eq_2}
	\end{equation}
	Based on (\ref{eq_1}), 
	the estimation error of plant state $\mathbf{e}_{k+1}=\mathbf{x}_{k+1}-\hat{\mathbf{x}}_{k+1}$  can be expressed as
	\begin{equation}
		{\mathbf{e}}_{k+1} = \begin{cases}
			\mathbf{w}_k, & \delta_{a,k}\xi_{a,k}=1, \\
			\mathbf{A}\mathbf{e}_{k}+\mathbf{w}_{k}, & otherwise.
		\end{cases}\label{eq_3}
	\end{equation}

	Then we denote the AoI of the sensor $S_a$ at time slot $k$ by  $\Delta_k$,  reflecting the information freshness of sensor $S_a$'s packets  at the controller. The  updating rule of AoI is as follows
	\begin{equation}
		\Delta_{k+1} = \begin{cases}
			1, &   \delta_{a,k}\xi_{a,k}=1, \\
			\Delta_k+1, &   otherwise,
		\end{cases}\label{eq_4}
	\end{equation}
	based on which the estimation error is further expressed as 
	\begin{equation}
		\mathbf{e}_{k+1}=\sum_{i=1}^{\Delta_{k+1}}\mathbf{A}^{i-1}\mathbf{w}_{k+1-i}.\label{eq_5}
	\end{equation}
	Then the covariance of estimation error is given by
	\begin{equation}
		\mathbf{\Theta}_{k+1}=\sum_{i=1}^{\Delta_{k+1}}\mathbf{A}^{i-1}\mathbf{R_w}(\mathbf{A}^{\top})^{i-1}.\label{eq_6}
	\end{equation}
	
	\subsubsection{Markov Process}
	Sensor $S_b$ monitors the Markov process-based environment  to obtain the context information. The context information-estimator at the controller is given as
	\begin{equation}
		\hat{v}_{k+1} = \begin{cases}
			v_k, &   \delta_{b,k}\xi_{b,k}=1, \\
			\hat{v}_k, &   otherwise.
		\end{cases}\label{eq_7}
	\end{equation}
	We denote $\Upsilon_{k+1}$ as the estimation quality of context information at the controller, \emph{i.e.}, $(v_{k+1}, \hat{v}_{k+1})$. Specifically, Let $\Upsilon_{k+1}=\left\{1,2,3,4\right\}$ represent the case of $(v_{k+1}=0, \hat{v}_{k+1}=0)$, $(v_{k+1}=0, \hat{v}_{k+1}=1)$, $(v_{k+1}=1, \hat{v}_{k+1}=0)$, $(v_{k+1}=1, \hat{v}_{k+1}=1)$, respectively.
	Then  the controller feeds the estimation results back to the scheduler and computes the control command 
	based on the received and estimated results.

	\subsection{Control Design and Plant State Analysis}	
	%			The control input of the one-step controllable case received by the actuator can be expressed as $u_{k+1}=\hat{u}_k$.
	The event-triggered controller starts to control when the plant state violates the preset value in a certain condition.
	As mentioned in Subsection II-B, we assume that the equipment may have different tolerances with different contexts (\emph{e.g.}, $\zeta_1$ for the case of $v=1$ and $\zeta_0$ for the case of $v=0$)  to the same plant state\cite{9382948}. 
	Then, the  control input is given by
	\begin{equation}
		\mathbf{u}_{k+1}=\begin{cases}
			\mathbf{G}\hat{\mathbf{x}}_{k+1}, & \delta_{c,k+1}=1, \\
			0, & otherwise,
		\end{cases}\label{eq_8}
	\end{equation}
	where $\mathbf{G}=-\frac{\mathbf{A}}{\mathbf{B}}$ represents the control gain\cite{8865111}. 
	$\delta_{c,k+1}$ is an indicator of reactive control at time slot $k+1$, given by
	\begin{equation}
		\delta_{c,k+1}=\begin{cases}
			\mathbb{I}\left\{\max\left\{0,|\mathbf{C}\hat{\mathbf{x}}_{k+1}|-\zeta_1\right\}>0\right\}, & \hat{v}_{k+1}=1, \\
			\mathbb{I}\left\{\max\left\{0,|\mathbf{C}\hat{\mathbf{x}}_{k+1}|-\zeta_0\right\}>0\right\}, & \hat{v}_{k+1}=0,
		\end{cases}\label{eq_9}
	\end{equation}
	where $\mathbb{I}(\cdot)$ is the indicator function.
	Based on the LTI model in (\ref{eq_1}) and control input in (\ref{eq_8}),
	the plant state is given by
	\begin{equation}
		\mathbf{x}_{k+2}=\begin{cases}
			\mathbf{A}(\mathbf{x}_{k+1}-\hat{\mathbf{x}}_{k+1})+\mathbf{w}_{k+1}, &  \delta_{c,k+1}\xi_{c,k+1}=1, \\
			\mathbf{A}\mathbf{x}_{k+1}+\mathbf{w}_{k+1}, & otherwise,
		\end{cases}\label{eq_10}
	\end{equation} 
	where $\xi_{c,k+1}$ is the indicator of successful  transmission for the controller-actuator link at time slot $k+1$.
	Thus, the plant-state covariance, $\mathbf{\Phi}_k=\mathbb{E}\left[\mathbf{x}_{k}\mathbf{x}_{k}^\top\right]$, has the following updating rule
	\begin{equation}
		\mathbf{\Phi}_{k+2}=\begin{cases}
			\mathbf{A}\mathbf{\Theta}_{k+1}\mathbf{A}^{\top}+\mathbf{R_w}, & \delta_{c,k+1}\xi_{c,k+1}=1, \\
			\mathbf{A}\mathbf{\Phi}_{k+1}\mathbf{A}^{\top}+\mathbf{R_w}, & otherwise.
		\end{cases}\label{eq_11}
	\end{equation}
	With successful downlink transmission,  the covariance of plant state is given by
	\begin{equation}
		\mathbb{E}[\mathbf{x}_{k+2}\mathbf{x}^{\top}_{k+2}]=\sum_{i=1}^{\Delta_{k+1}}\mathbf{A}^{i}\mathbf{R}_w(\mathbf{A}^{\top})^i+\mathbf{R}_w,\label{eq_12}
	\end{equation}
	which  can be expressed as a non-decreasing function of sensor's AoI $\Delta_k$.

	\subsection{Goal of the WNCS}
	We aim to minimize the impact of extreme plant state on the system performance and ensure the stability of WNCS. 
	Particularly, we consider the impacts of goal-related state and context information on  the system performance simultaneously.
	Then we formulate an objective function to capture  violation events and reflect the significance  of the goal-related	state  at the point of actuation with different contexts, given by
	\begin{equation}
		J_{k+2}=\begin{cases}
			\mathbb{I}\left\{\max\left\{0,|\mathbf{C}{\mathbf{x}}_{k+2}|-\zeta_1\right\}>0\right\}, & {v}_{k+2}=1, \\
			\mathbb{I}\left\{\max\left\{0,|\mathbf{C}{\mathbf{x}}_{k+2}|-\zeta_0\right\}>0\right\}, & {v}_{k+2}=0.
		\end{cases}\label{eq_13}
	\end{equation}
	\textit{This implies different significance of state information and context information,  in relation to the goal.}	
	
	According to (\ref{eq_10}) and (\ref{eq_12}), it can be seen that frequent updating of  sensor $S_a$ and triggering of control action contributes to the reduction of the  goal-related state, consequently reducing the associated violation probability.
	Moreover, accurate estimation of context information helps to avoid the occurrence of violations and excessive controls.
	Therefore, to minimize the average  violation probability with the given cost, we formulate the following problem to jointly schedule multiple heterogeneous packets and design control inputs.
	\begin{subequations}
		\begin{align}
			\mathcal{P} 1: 	\min_{\Pi,\mathbf{U}} \quad  & J=\lim\limits_{K\rightarrow\infty}\frac{1}{K}\sum_{k=1}^{K}{\mathbb{E}\left\{J_k\right\}}\\
			\textrm{s.t.} \quad & \delta_{a,k}+\delta_{b,k}\leq1, k\in\left[1,K\right],\\
			&\lim\limits_{k\rightarrow\infty}\frac{1}{K}\sum_{k=1}^{K}\mathbb{E}\left\{c_k\right\}\leq\overline{c}_{\max}, \label{eq_14}
		\end{align}
	\end{subequations}where ${\Pi}=[\pi_1,\pi_2,\cdots,\pi_K]$ denotes the scheduling policy with ${\pi_k}=(\delta_{a,k}, \delta_{b,k})$ at time slot $k$.
	$\mathbf{U}=[\mathbf{u}_1,\mathbf{u}_2,\cdots,\mathbf{u}_K]$ denotes the control input at each time slot.
	{$c_k=c_a\delta_{a,k}+c_b\delta_{b,k}+c_c\delta_{c,k}$ is the total cost, where  $c_a$, $c_a$ and $c_c$ are the action cost for sensors $S_a$, $S_b$ and the controller, respectively.}
	Recall that $\delta_{a,k}$ and $\delta_{b,k}$ represent the indicators of updating actions for  sensors $S_a$ and $S_b$, respectively. Also, $\delta_{c,k}=1$ indicates that the controller is triggered at time slot $k$.
	(14b) shows that at most one updating action is allowed in each time slot. (14c) limits the maximum allowed total cost 
	$\overline{c}_{\max}$, including both the updating and control costs.

	\section{Goal-oriented Deterministic Scheduling Policy}
	
	In this section, we formulate $\mathcal{P}$1 as a CMDP and reformulate it as an MDP by using Lagrangian relaxation method. Then we propose a deterministic scheduling policy based on samples and feedback estimation results at the scheduler.

	\subsection{CMDP Formulation}

	Recall that $\pi$ denotes the scheduling policy that determines the action taken at each state. A stationary randomized policy is mapping from each state to a distribution over actions. 
	Let $\bar{J}^\pi=\lim\limits_{K\rightarrow\infty}\frac{1}{K}\sum_{k=1}^{K}{\mathbb{E}_{\pi}\left\{J_k\right\}}$ denote the average  violation probability with the scheduling policy $\pi$. Let $\bar{c}^\pi=\lim\limits_{k\rightarrow\infty}\frac{1}{K}\sum_{k=1}^{K}\mathbb{E}_{\pi}\left\{\tilde{c}_k\right\}$ denote the average number updating actions, where $\tilde{c}_k=c_a\delta_{a,k}+c_b\delta_{b,k}$ is the scheduling cost. The cost is simplified first due to that we focus on the scheduling policy in this section and   the control cost depends on the control policy. By integrating the constraint (14b) into the scheduling policy $\pi$, problem (\ref{eq_14}) is equivalently cast as the CMDP
	\begin{subequations}
		\begin{align}
			\mathcal{P} 2: 	\min_{\pi} \quad  & \bar{J}^\pi\\
			\textrm{s.t.} \quad & \bar{c}^\pi\leq\overline{c}_{\max}. 
		\end{align}\label{eq_15}
	\end{subequations}
	A typical CMDP  consists of the state space $\mathcal{S}$, action space $\mathcal{A}$, state transition matrix $\mathcal{T}$ and reward $\mathcal{R}$, defined as
	\begin{itemize}
		\item 
		\textbf{State:} The state space contains two states and is defined as $\mathcal{S}=\left\{\Delta, \Upsilon\right\}$. $\Delta$ is the set of AoI for sensor $S_a$, which {reflects the value of goal-related state} and is defined as the time elapsed since the last successfully received sensor’s packet at the receiver. The plant state of the WNCS is a non-decreasing function of sensor $S_a$'s AoI, as analyzed in Subsection III-B.				
		$\Upsilon=\left\{1,2,3,4\right\}$, monitored by sensor $S_b$, is the set of estimation quality indicator of context, which also affects the value of WNCS's goal. 
		The state of the CMDP at time $k$ is $s_k=(\Delta_k,\Upsilon_k)\in\mathcal{S}$.
		
		\item 
		\textbf{Action:} The action space of the CMDP is defined as $\mathcal{A}=\left\{1,2,3\right\}$. With the deterministic scheduling policy $\pi$, the action at time $k$, \emph{i.e.}, {$\alpha_k=\pi(s_k)\in\mathcal{A}$}, indicates the sensor $S_a$'s transmission ($\alpha_k=1$) or the sensor $S_b$'s transmission ($\alpha_k=2$)  or being idle ($\alpha_k=3$).

		\item 
		\textbf{Transition matrix:} The state-transition probability $P(s'|s,\alpha)$ is the probability that the state $s$ at time slot $k-1$ transits to $s'$ at time $k$ with action $\alpha$ at time $k-1$.
		By omitting the subscript $k$, let $s=(\Delta,\Upsilon=1)$ and $s'$ denote the current and next state, respectively.
		The state-transition probability can be obtained as follows.
		{\small
			\begin{equation}
				P(s'|s,\alpha)=\begin{cases}
					\epsilon_a\bar{p}, & if\quad \alpha=1, s'=(\Delta+1,\Upsilon=1) \\
					\epsilon_ap, & if\quad \alpha=1, s'=(\Delta+1,\Upsilon=3)  \\
					(1-\epsilon_a)\bar{p}, & if\quad \alpha=1, s'=(1,\Upsilon=1) \\
					(1-\epsilon_a)p, & if\quad \alpha=1, s'=(1,\Upsilon=3) \\
					\bar{p}, & if\quad \alpha=2, s'=(\Delta+1,\Upsilon=1) \\
					p, & if\quad \alpha=2, s'=(\Delta+1,\Upsilon=3) \\
					%							(1-\xi_b)P_{i,i}, & if\quad a=2, s'=(\Delta+1,\Upsilon=1) \\
					%							(1-\xi_b)P_{i,j}, & if\quad a=2, s'=(\Delta+1,\Upsilon=1)\\
					\bar{p}, & if\quad \alpha=3, s'=(\Delta+1,\Upsilon=1) \\
					{p}, & if\quad \alpha=3, s'=(\Delta+1,\Upsilon=3) \\	 
					0, &otherwise.
				\end{cases}\label{eq_16}
		\end{equation}}Similarly, the state-transition probability for the other initial cases, \emph{i.e.},  $s=(\Delta,\Upsilon=2,3,4)$, can also be derived, as shown in Appendix A.
		
		\item 
		\textbf{Reward function:}  $\mathcal{R}_k$
		is the instantaneous reward when	it takes action $\alpha_k$ in state $s_k$, which maps a state-action pair to a real number. The CMDP has two cost functions: 1) scheduling cost, defined as $c(\alpha_k)=\tilde{c}_k$, \emph{i.e.}, $c(\alpha_k)=c_a$ or $c(\alpha_k)=c_b$ if the scheduler makes a transmission attempt at slot $k$, otherwise $c(\alpha_k)=0$, and 2) violation cost, defined as $\chi(s_k)=J_k$, \emph{i.e.}, $\chi(s_k)=1$ if the current state violates the threshold at slot $k$, otherwise $\chi(s_k)=0$.
		%				 Combined with our optimization problem, the reward function
		%				can be defined and will be analyzed in the following subsection.	
	\end{itemize}

	\subsection{Problem Reformulation and Analysis}

	Since the problem in (15) is a CMDP, which is, in general, difficult
	to be solved\cite{2023arXiv230304908F}, we   utilize the Lagrangian relaxation method to transform
	it to an unconstrained MDP\cite{9137714}. Then the analysis of state space and reward function for the formulated MDP is provided. 
	\subsubsection{MDP formulation}
	The Lagrangian relaxation method is adopted to transform  the CMDP to an  MDP\cite{9137714}.
	The Lagrangian function is defined as
	\begin{small}
		\begin{equation}
			\mathcal{L}(\pi,\lambda)=\lim\limits_{K\rightarrow\infty}\frac{1}{K}\sum_{k=1}^{K}\mathbb{E}_{\pi}\left\{J_k+\lambda \tilde{c}_k\right\}-\lambda \overline{c}_{\max},\label{eq_17}
		\end{equation}
	\end{small}where $\lambda$ is the Lagrangian multiplier and the immediate cost is $\mathcal{R}_k=J_k+\lambda \tilde{c}_k$.
	Then
	$\mathcal{P} 2$ can be transformed as 	$\min_{\pi\in\Pi}	\mathcal{L}(\pi,\lambda)$,
	for any given $\lambda\ge0$.	Since $\lambda \overline{c}_{\max}$ is independent of the chosen policy $\pi$, 	$\mathcal{P} 2$ is equivalent to solving the following optimization problem
	\begin{small}
		\begin{equation}
			\mathcal{P} 3:	\min_{\pi\in\Pi}h(\lambda,\pi)=\min_{\pi\in\Pi}\limsup_{K\rightarrow\infty}\frac{1}{K}\sum_{k=1}^{K}\mathbb{E}_{\pi}\left\{J_k+\lambda \tilde{c}_k\right\}.\label{eq_18}
		\end{equation}
	\end{small}
	
	Let $\pi_\lambda^*$ denote an optimal policy that solves problem (18) for a given $\lambda$, which is called a $\lambda$-optimal policy.
	Then we provide the problem analysis for finding the optimal scheduling policy in the following subsections.
	
	\subsubsection{Truncated State Space}

	Since the formulated  MDP has an infinite number of state $\Delta$, we approximate it by
	a truncated MDP with finite states.
	Based on (12) and (13), we can conclude that  $J_k$ is a bounded increasing function of state $\Delta$. 
	It implies that when the state $\Delta$ is larger than a certain threshold, $J_k$ converges to 1 and remains constant.
	With the state threshold $\Delta_{\text{thr}}$, the set of AoI is denoted by $\Delta=\left\{1,2,\cdots,\Delta_{\text{thr}}\right\}$. Then we have $J_k=1$ if $\Delta_{k}\ge\Delta_{\text{thr}}$, and $J_k=0$ if $\Delta_{k}<\Delta_{\text{thr}}$, as provided in Proposition 1.
	%, where for any $\Delta_k>\Delta_{\text{thr}}$, we have $f_v(k)=1$.
	
	\textbf{Proposition 1:}  With the given LTI system, the state space of the MDP can be truncated and the state threshold is derived as $\Delta_{\text{thr}}=\underset{v=\left\{0,1\right\}}{\max}\left\{ \arg\min\left\{\Delta\in\mathbb{N}^+|\mathbf{C}\sqrt{\mathbf{\Phi}(\Delta)}>\zeta_v \right\}\right\}$.
	
	\textit{Proof: 	Based on (12),  the expectation of plant state can be calculated as $\mathbb{E}[|\mathbf{x}_k|]=\sqrt{\mathbf{\Phi}_k}$, and the plant-state covariance is a non-decreasing function of AoI. Hence, finding the minimal value of $\Delta$ that satisfies $\mathbf{C}|\mathbf{x}_k|>\zeta_v$ is equivalent to solving the following equation, $\Delta_{v,\text{thr}}=\arg\min\left\{\Delta\in\mathbb{N}^+|\mathbf{C}\sqrt{\mathbf{\Phi}(\Delta)}>\zeta_v \right\}$. By comparing $\Delta_{0,\text{thr}}$ and $\Delta_{1,\text{thr}}$,  the state threshold $\Delta_{\text{thr}}=\max\left\{\Delta_{0,\text{thr}},\Delta_{1,\text{thr}}\right\}$ can be derived. }	 $\hfill\blacksquare$

	\subsubsection{Reward Function Mapping}
	
	%		Without loss of generality and for ease of analysis, we consider a case that each action consumes  a unit cost, \emph{i.e.}, $c_a=c_b=1$. 
	Now we turn to calculating the reward mapping function in relation to the state and action.
	As shown in Subsection IV-A, the instantaneous reward function consists of scheduling cost and violation cost.
	The scheduling cost can be obtained directly based on the actions, \emph{i.e.}, $c(\alpha_k)=c_a$ or $c(\alpha_k)=c_b$, if the scheduler makes a transmission attempt at slot $k$. The violation cost can be calculated based on (\ref{eq_12}) and (\ref{eq_13}), and then we have Definition 1 to calculate the reward function.
	
	\textbf{Definition 1:}  	The mapping function for the violation cost and state, $\chi(\Delta_k,\Upsilon_k)=J_k$,  reflects whether the current state results in a violation, which can be expressed as 
	\[
	\chi(\Delta,\Upsilon)=
	\Delta\left\{\overbrace{\begin{bmatrix}
			0 & 0 & 0 & 0 \\
			%	\vdots & \vdots & \vdots & \vdots \\
			\vdots & \vdots & 1 & 0 \\
			0 & 0 & 1 & 1 \\
			\vdots & \vdots & \vdots & \vdots \\
			1 & 1 & 1 & 1
	\end{bmatrix}}^{\Upsilon=1,2,3,4}.\right.\\\tag{19}
	\]
	
	\textit{Proof: When the current state satisfies $\Delta_k<\Delta_{0,\text{thr}}$, $\chi(\Delta_k,\Upsilon_k)=0$ holds regardless of the transmission and is independent of $\Upsilon_k$. In the case of $\Delta_k\ge\Delta_{1,\text{thr}}$,  $\chi(\Delta_k,\Upsilon_k)=1$ for all the contexts.
		When $\Delta_{0,\text{thr}}<\Delta_k<\Delta_{1,\text{thr}}$, $\chi(\Delta_k,\Upsilon_k)=1$ for the case of $\Upsilon_k=3$ and $\Upsilon_k=4$.
		Also, due to the missing control command, $\chi(\Delta_k,\Upsilon_k)=1$ holds for the case of  $\Delta_k=\Delta_{0,\text{thr}}-1$ with $\Upsilon_k=3$.
		%			In other cases, to track plant state and avoid overcontrol, $\chi(\Delta_k,\Upsilon_k)=0$ holds for action $a=1$ when $\Upsilon_k=2$.  To update context information and avoid missing control, $\chi(\Delta_k,\Upsilon_k)=0$ holds for action $a=2$ when $\Upsilon_k=3$.
	} $\hfill\blacksquare$
	
	Hence, the reward function can be designed by mapping state-action pairs to  real numbers	  with given states and actions.
	{The violation cost $\chi(\Delta,\Upsilon)$ can be reshaped as a matrix $\mathbf{R}_s$ of $4\Delta_{\text{thr}}\times1$. 
		Considering the scheduling cost $\tilde{c}_k$ and Lagrangian multiplier $\lambda$, the reward function is expressed as
		$\mathbf{{R}}=\left[-\mathbf{R}_s-\lambda c_a, -\mathbf{R}_s-\lambda c_b, -\mathbf{R}_s \right]$.}
	
	\subsection{Deterministic Scheduling Policy}
	\vspace{-2pt}
	This subsection proposes the deterministic scheduling policy based on the above analysis.
	By ensuring $\max\left\{\epsilon_a,\epsilon_b,\epsilon_c\right\}<\frac{1}{\rho^2(\mathbf{A})}$, there exists a
	stationary and deterministic optimal scheduling policy that can stabilize the  controllable  plant\cite{8865111}.			Then the challenging problem $\mathcal{P} 3$ can be solved by  using the RVIA and bisection method iteratively \cite{2023arXiv230304908F,8865111}.

	\vspace{3pt}
	\subsubsection{RVIA for $\lambda$-Optimal Policy}
	
	To obtain an optimal policy $\pi_\lambda^*$ for a given $\lambda$, we solve the MDP via RVIA. There exists a relative value function $V(s)$, $s\in\mathcal{S}$, that satisfies 
	\setcounter{equation}{19}
	\begin{small}
		\begin{equation}
			\bar{L}^*(\lambda)+V(s)=\min_{\alpha\in\mathcal{A}_s}\left[L(s,\alpha,\lambda)+\sum_{s'\in\mathcal{S}}Pr(s'|s,\alpha)V(s')\right],\label{eq_20}
		\end{equation}
	\end{small}where	$\bar{L}^*(\lambda)$ is the optimal value of the MDP problem for a given $\lambda$, defined as 	$\bar{L}^*(\lambda)=\min_{\pi\in\Pi}\bar{L}(\pi,\lambda)$. Subsequently, the $\lambda$-optimal policy is obtained as
	\begin{small}
		\begin{equation}
			\pi_{\lambda}^*(s)=\arg\min_{\alpha\in\mathcal{A}_s}\left[L(s,\alpha,\lambda)+\sum_{s'\in\mathcal{S}}Pr(s'|s,\alpha)V(s')\right],\label{eq_21}
	\end{equation} \end{small}where the relative value function is updated as $V^i(s)=z^i(s)-z^i(s^{\text{ref}})$.  Note that $s^{\text{ref}}$ is an arbitrary reference state which remains unchanged throughout the iterations, and we have $\bar{L}(\lambda)=z(s^{\text{ref}})$. The term $z^i(s)$, called value function, is obtained at each iteration as
	\begin{small}
		\begin{equation}
			z^i(s)=\min_{a\in\mathcal{A}_s}\left[L(s,a,\lambda)+\sum_{s'\in\mathcal{S}}Pr(s'|s,a)V(s')\right].\label{eq_22}
		\end{equation}
	\end{small}

	For a given $\lambda$-optimal policy ($\pi^*_\lambda$), the objective function of the CMDP problem, $\bar{J}^{\pi^*_\lambda}$, and the objective function of the MDP problem, $\bar{L}^*(\lambda)$, are increasing with respect to $\lambda$. The constraint of the CMDP problem, $\bar{\delta}^{\pi^*_\lambda}$ is decreasing in relation to $\lambda$.

	\subsubsection{Bisection Search for $\lambda$}
	To determine $\tilde{\lambda}$, the bisection algorithm is adopted by utilizing the monotonicity of $\bar{\delta}^{\pi^*_\lambda}$ with respect to $\lambda$.
	We initialize the bisection algorithm with $\lambda_u$ and $\lambda_l$ in a manner that ensures $\bar{\delta}^{\pi^*_{\lambda_u}}\leq\overline{c}_{\max}$ and 
	$\bar{\delta}^{\pi^*_{\lambda_l}}\geq\overline{c}_{\max}$.
	Upon the termination of the bisection algorithm, \emph{i.e.}, the gap between $\lambda_{u}$ and $\lambda_{l}$ is smaller than the threshold $\kappa$, we set $\tilde{\lambda}=\lambda_u$ and obtain the best feasible $\lambda$-optimal policy as $\pi^*_{\tilde{\lambda}}=\pi^*_{\lambda_u}$. Additionally, the algorithm returns the infeasible policy associated with $\lambda_l$, which serves as a lower-bound to an optimal solution.

	Subsequently, at each iteration of the process,  a $\lambda$-optimal policy is developed for a given $\lambda$ via the RVIA. Following this, $\lambda$ is updated according to the bisection rule. The iterative procedure continues until the best $\lambda$-optimal policy among the feasible  policies is found, as summarized in Algorithm 1.

	\begin{algorithm}[!ht]
		\DontPrintSemicolon
		\small
		\caption{Deterministic scheduling policy }
		\textbf{Input:}  Cost constraint: $\overline{c}_{\max}$,  RVIA parameters: $s^{\text{ref}}$, $\iota$, Bisection algorithm parameters: $\lambda_u$, $\lambda_l$, $\kappa$.\\
		%\tcc{Bisection algorithm}		
		\While{$\lambda_u-\lambda_l\ge\kappa$}
		{
			$\bar{\lambda}=\frac{\lambda_u+\lambda_l}{2}$
			%\tcc{RVIA for the given $\lambda$}	
			
			\textbf{Initialize:}  $i=1$,  $V^0(s)=1$, $V^1(s)=0$, $z^0(s)=0$ for all $s\in\mathcal{S}$\\
			\While{$\max_{s\in\mathcal{S}}|V^i(s)-V^{i-1}(s)|\ge\iota$}
			{$i=i+1$\\
				\For{$s\in\mathcal{S}$}    
				{ 
					Perform	(21) and
					$V^i(s)=z^i(s)-z^i(s^\text{ref})$
				}
			}
			%\tcc{An optimal policy for given $\bar{\lambda}$}	
			\For{$s\in\mathcal{S}$}    
			{ 
				Perform (20)
			}
			\eIf{$\bar{\delta}^{\pi^*_{\bar{\lambda}}}\geq\overline{c}_{\max}$}
			{
				$\lambda_u$=$\bar{\lambda}$
			}
			{
				$\lambda_l$=$\bar{\lambda}$
			}
		}
		\textbf{Output:} Lagrangian multiplier: $\tilde{\lambda}=\lambda_u$, feasible policy: $\pi^*_{\bar{\lambda}}=\pi^*_{\lambda_u}$
	\end{algorithm}
	
	\vspace{-10pt}
	\subsection{Complexity Analysis} 
	The computational complexity of the proposed algorithm depends on the number of interactions and complexity of each iteration $\mathcal{O}(|\mathcal{A}||\mathcal{S}|^2)$. The state space size $|\mathcal{S}|$ is approximately $\Delta_{\text{thr}}\Upsilon_{\text{thr}}$ and the action space size
	$|\mathcal{A}|$ is $3$. Accordingly, the complexity of the deterministic policy
	is $\mathcal{O}(3I_1I_2\Delta_{\text{thr}}^2\Upsilon_{\text{thr}}^2)$, where $I_1$ and $I_2$, are, respectively, the iterations required in bisection and RVIA.

	\section{Goal-oriented Estimation, Communication and Control Co-Design}
	In this section, we investigate the goal-oriented co-design of estimation, communication, and control based on the scheduling policy proposed in the last section. Specifically, we propose control-aware estimation,  sensing-assisted control, and state-dependent retransmission strategies to attain a lower violation probability with the same cost.

	\subsection{Goal-Oriented Scheduling and Control}
\begin{figure}[htbp]
		\centering
		{\includegraphics[height=4.2cm]{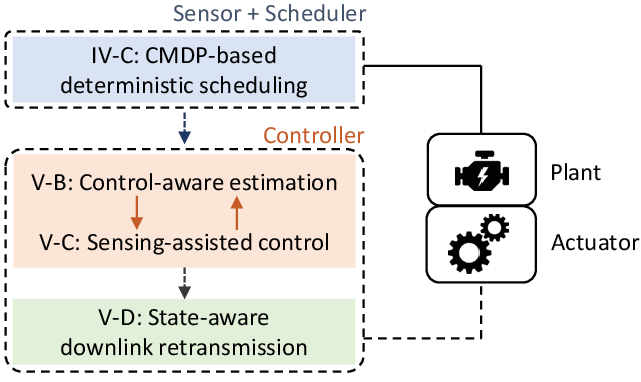}}
		\caption{Flow chart of the improved GSC strategies.}\label{fig_3}
	\end{figure}

	Based on the deterministic scheduling policy proposed in Section IV-C, as well as the control input in Subsection III-B, we propose the goal-oriented scheduling and control  (GSC) co-design strategy. The scheduler and controller are designed in sequence, thanks to their independence\cite{8845114}.  To further reduce the violation probability and cost, more advanced communication and control policies are introduced in the following subsections. The flow chart of the improved GSC scheme is shown in Fig. \ref{fig_3}.
	
	\subsection{Control-Aware Sensing/Estimation}
	This subsection revises the state estimation process based on the control command. In subsection III-A, the plant state is estimated based on the assumption that the controller is triggered all the time, \emph{i.e.}, $\mathbf{u}=\mathbf{-\frac{A}{B}}\hat{\mathbf{x}}$, as shown in (\ref{eq_2}) and (\ref{eq_8}). 
	Hence,  the estimated plant state without control information can be expressed as 
	\begin{equation}
		\hat{\mathbf{x}}_{k+1} = \begin{cases}
			\mathbf{A}{\mathbf{e}}_{k}, & \delta_{a,k}\xi_{a,k}=1, \\
			\textbf{0}, & otherwise.
		\end{cases}\label{eq_23}
	\end{equation}
	
	\textbf{Control-aware estimation (CAE)}: Due to that the controller is event-triggered, the control command is zero when the estimated plant state is smaller than the state threshold. 
	Hence, we propose the CAE scheme based on the control information, as shown in Proposition 2.
	
	\textbf{Proposition 2:}  
	By utilizing the information on whether the controller is triggered or not, the  CAE scheme is able to improve the accuracy of the estimated state of the plant, given by
	\begin{equation}
		\hat{\mathbf{x}}_{k+1} = \begin{cases}
			\mathbf{A}{\mathbf{e}}_{k}, & \delta_{a,k}\xi_{a,k}=1,  \delta_{c,k+1}=1,\\
			\mathbf{A}{\mathbf{x}}_{k}, & \delta_{a,k}\xi_{a,k}=1,  \delta_{c,k+1}=0,\\
			\textbf{0}, & \delta_{a,k}\xi_{a,k}=0,   \delta_{c,k+1}=1,\\
			\mathbf{A}\hat{\mathbf{x}}_{k}, & \delta_{a,k}\xi_{a,k}=0,  \delta_{c,k+1}=0. 
		\end{cases}\label{eq_24}
	\end{equation}	 
	
	\textit{Proof:  With the control input $\mathbf{u}=\mathbf{-\frac{A}{B}}\hat{\mathbf{x}}$, the estimated plant state at $k+1$ slot based on the successful/unsuccessful uplink transmission of sensor $S_a$ at  $k$ slot  is given by $\hat{\mathbf{x}}_{k+1}=	\mathbf{A}{\mathbf{e}}_{k}$ and $\hat{\mathbf{x}}_{k+1}=\textbf{0}$, respectively.
		Moreover, the plant state  is  respectively estimated as $\hat{\mathbf{x}}_{k+1}=	\mathbf{A}{\mathbf{x}}_{k}$ and $\hat{\mathbf{x}}_{k+1}=	\mathbf{A}{\mathbf{\hat{x}}}_{k}$ without control input.}	  $\hfill\blacksquare$
	
	\textbf{Remark 1:}	By comparing (\ref{eq_23}) and (\ref{eq_24}), it is shown that the CAE-based plant state is more accurate by considering the case of no control inputs, based on which a more reliable control input can be updated based on (\ref{eq_8}).

	\subsection{Sensing-Assisted Control}
	%			In this subsection, we improve the control strategy by utilizing the 
	The reactive control introduced in Subsection III-B has an inevitable drawback, that is the controller only reacts when the state has already violated the threshold, leading to a high violation probability.
	To address this problem, we propose two advanced control strategies to further improve the performance in this subsection.

	\textbf{Proactive control (PC)}: The PC strategy is proposed to limit the plant state proactively,  by predicting the future plant state at the next time slot without control input. In this case, the controller can transmit the control command in advance before the plant state violates the threshold. 
	The control rule is given as 
	\begin{equation}
		\delta_{c,k+1}=\begin{cases}
			\mathbb{I}\left\{\max\left\{0,|C\hat{\mathbf{x}}_{k+2}|-\zeta_1\right\}>0\right\}, & \hat{v}_{k+2}=1, \\
			\mathbb{I}\left\{\max\left\{0,|C\hat{\mathbf{x}}_{k+2}|-\zeta_0\right\}>0\right\}, & \hat{v}_{k+2}=0,
		\end{cases}\label{eq_25}
	\end{equation}
	where
	$\hat{\mathbf{x}}_{k+2} = \mathbf{A}{\mathbf{\hat{x}}}_{k+1}$ is the future plant state without control.
	
	Since $\rho(\mathbf{A})>1$, the control rule proposed in (24)  implies that the PC strategy is able to reduce the violation probability by inducing more control actions. 
	However, the performance of the PC strategy depends on the accuracy of estimated plant state and contextual information at the next time slot, which may introduce more errors. 	
	
	\textbf{Conservative reactive control (CRC)}:	The proposed CRC strategy uses the conservative ratio to bring down the threshold to avoid violations.
	In this case, the plant state triggers the control action before  violating the preset threshold.
	The triggered condition for the controller in (9) can be revised as
	\begin{equation}
		\delta_{c,k+1}=\begin{cases}
			\mathbb{I}\left\{\max\left\{0,|C\hat{\mathbf{x}}_{k+1}|-\theta\zeta_1\right\}>0\right\}, & \hat{v}_{k+1}=1, \\
			\mathbb{I}\left\{\max\left\{0,|C\hat{\mathbf{x}}_{k+1}|-\theta\zeta_0\right\}>0\right\}, & \hat{v}_{k+1}=0,
		\end{cases}\label{eq_26}
	\end{equation}
	where $0<\theta<1$ is the conservative ratio.
	
	It is shown that the CRC strategy depends on the selection of conservative ratio. A smaller $\theta$ results in a lower violation probability, but may lead to overcontrol and induce higher control cost.

	\subsection{State-Dependent Downlink Retransmission}
	In this subsection, we propose the {state-dependent retransmission strategy} to ensure reliable control, which may lead a low violation probability for WNCSs.
	As defined in Subsection II-C,  the probability of unsuccessful downlink transmission is given by ${\Pr}\left[\xi_{c,k}=0|\delta_{c,k}=1\right]=1-\epsilon_c$, where  $\delta_{c,k}$ and $\xi_{c,k}$ respectively represent the indicator of triggered control and successful downlink transmission at time slot $k$.

	\textbf{Truncated automatic repeat request (TARQ):} 
	Recall that the control input at the actuator is zero when the downlink transmission error occurs. This leads to unstable control and high violation probability when the error probability is high. Hence, we adopt the TARQ transmission scheme with a limited number of maximum allowable retransmissions $n_{\max}$, which improves the downlink transmission reliability by retransmitting the same packets, as shown in Fig. 4. 
	\vspace{-7pt}
	\begin{figure}[htbp]
		\centering
		{\includegraphics[height=2cm]{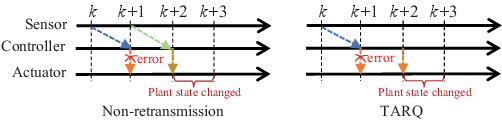}}
		\caption{Flow chart of the improved GSC strategies.}\label{fig_3_add}
	\end{figure}
	
	The control input with $n-1, n\in \left[1,\cdots,n_{\max}\right]$ retransmissions is calculated as $\mathbf{u}_{k+n}=-\mathbf{\frac{A}{B}}\hat{\mathbf{x}}_{k+1}$. 
	Then the plant state is given by 
	\begin{small}
		\begin{equation}
			\mathbf{x}^{\text{ARQ}}_{k+1+n}=\begin{cases}
				\mathbf{A}(\mathbf{x}_{k+n}-\hat{\mathbf{x}}_{k+1})+\mathbf{w}_{k+n}, &  \omega_{c,k+n}=1, \\
				\mathbf{A}\mathbf{x}_{k+n}+\mathbf{w}_{k+n}, & otherwise.
			\end{cases}\label{eq_27}
		\end{equation} 
	\end{small}where $\omega_{c,k+n}=\delta_{c,k+n}\xi_{c,k+n}$ is the indicator of downlink transmission.
	{To further reduce the  cost, we assume that the scheduler stops transmitting sensor states during downlink retransmission.}
	However, retransmitting the historical control commands fails to capture plant state change and leads to inappropriate  operations.
	Hence, retransmission is not always beneficial in reducing the violation probability.

	%	The TARQ scheme is suitable for the case of high error probability, while the non-retransmission scheme achieves a lower plant state for the dynamic plant with low error probability.
	\textbf{Proposition 2:}  
	With $n-1$ retransmissions, the difference between the plant states obtained by TARQ and non-retransmission schemes is given by 
	\begin{small}
		\begin{equation}
			\begin{aligned}
				\mathbb{E}\left[\mathbf{D}\right]=&	\mathbb{E}\left[\mathbf{\Phi}_{k+1+n}\right]^{\text{ARQ}}-\mathbb{E}\left[\mathbf{\Phi}_{k+1+n}\right]^{\text{Non}}\\
				=& (\epsilon-\epsilon^{n})(\mathbf{A}\mathbf{\Theta}_{k+n}\mathbf{A}^{\top}-\mathbf{A}\mathbf{\Phi}_{k+n}\mathbf{A}^{\top})\\
				&+(1-\epsilon^{n})\left((\mathbf{A}^{n+1}-\mathbf{A}^2)\mathbf{\Phi}_{k}(\mathbf{A}^{n+1}-\mathbf{A}^2)^{\top}\right),
			\end{aligned}\label{eq_28}
		\end{equation}	
	\end{small}which is  decreasing with  respect to the error probability for the quasi-static plant but increases  for the dynamic plant.

	\textit{Proof:	Please refer to Appendix B.} $\hfill\blacksquare$
	
	Proposition 2 implies that  the optimal transmission policy depends on the error probability and the system matrix.

	\section{Case Study and Simulation Results}
	In this section, we consider the LFC system in smart grid as a case study. Also, we 
	compare them with the existing work in terms of violation probability and cost.

	\vspace{-10pt}
	\subsection{Parameters Setting}
	The goal of  the considered LFC system is to maintain the balance between power load and generation at minimal cost\cite{9807392}.
	However, cyber-attacks can affect the frequency regulation, leading to power system instability.
	{Sensors $S_a$ and $S_b$ are employed to  monitor the frequency deviation and detect any cyberattack, respectively\cite{9382948}.}
	%Note that several detection algorithms have been developed to detect  attacks in power systems. 
	In this paper, we focus on data transmission and assume that the sensor $S_b$ has the ability to detect cyberattacks\cite{TSG2949998}.  
	Based on the received packets that convey frequency deviation and context information, the controller guides the generator to minimize the frequency deviation.
	The grid may be unstable if the frequency deviation violates a given threshold, which  depends on context information, \emph{i.e.}, whether there is a cyberattack. 
	
	To verify the effectiveness of the proposed strategies,		
	the following methods are compared as benchmarks.
	
	\begin{enumerate}
		\item 
		\textit{Round-robin (RR) scheduling\cite{8995639}:} At each time slot, the scheduler periodically switches between sensor $S_a$ and sensor $S_b$ for updating. 
		\item 
		\textit{{Randomized  selection} (RS) scheduling\cite{9162973}:} At each time slot, the scheduler randomly selects sensor $S_a$ or sensor $S_b$ for updating.
		\item 		
		\textit{AoI-aware scheduling\cite{8845114}:} The scheduler selects one of the sensors when the sensor's AoI violates the preset value while ignoring contextual information. 
		%The AoI threshold is set to 3. 
		% Otherwise, it remains idle. 
		\item 
		\textit{AoII-aware scheduling\cite{9137714}:} The scheduler selects the sensor $S_b$ when  incorrect DTMC status estimating occurs while ignoring  the goal. Otherwise the scheduler selects the sensor $S_a$ or remains idle based on its AoI.
		%	\textbf{Time-triggered control:} The methods that transmit control command at each time slot are also provided for comparison.	
	\end{enumerate}

	Simulations are carried out via Monte Carlo simulations with 1,000,000 generated	packets. 
	To compare different algorithms, violation probability and normalized cost are adopted to indicate  plant stability and its resource consumption.
	The violation probability is calculated by the probability of a violation occurring in all samples. Since we consider different action costs,  the normalized total cost is the weighted sum of each action divided by the maximum cost, given by $(\delta_a*c_a+\delta_a*c_b+\delta_c*c_c)/(K*c_a+K*c_c)$. 
	{Also, 10 KHz bandwidth is allocated for sensing  updates with the packet size of 2 Kbits, using quadrature phase shift keying modulation\cite{3GPP_service}. For consistency in communication, control and plant evolution, one time slot is normalized to 1 second.}
	Parameters are provided in Table II unless otherwise noted.	
	
	\vspace{-5pt}
	\renewcommand{\arraystretch}{1.1}
	\begin{table}[h]
		\centering
		\caption{Parameter setting}
		\footnotesize
		\begin{tabular}{lllll}
			\cline{1-2}
			Parameter                                & Value &  &  &  \\ 	\cline{1-2}
			System matrix               & $\mathbf{A}=
			\left[ {\begin{array}{ccc}
					-0.08 & 6 & 0\\
					0&-0.25 & 0.25\\
					-0.4167&0&-1.25\\
			\end{array} } \right]$  &  &  &  \\ 	\cline{1-2}
			Control matrix  & $\mathbf{B}=\left[0, 0, 1.25\right]^{\top}$     &  &  &  \\ \cline{1-2}
			Measurement matrix  & $\mathbf{C}=\left[1, 0, 0\right]^{\top}$     &  &  &  \\ 	\cline{1-2}
			Transition  matrix  & $\mathbf{P}=
			\left[ 0.8, 0.2; 0.2 , 0.8\right]$     &  &  &  \\ 	\cline{1-2}
			%			Sampling period                          & $T_s=1$ s    &  &  &  \\ 	\cline{1-2}
			%			Number of samples                        & $K=1000000$     &  &  &  \\ 	\cline{1-2}
			Process noise variance                   & $\mathbf{R_w}=1e-7\mathbf{I}$     &  &  &  \\ 	\cline{1-2}
			Error probability                        &  $\epsilon_a=\epsilon_b=\epsilon_c=1e-3$    &  &  &  \\ 	\cline{1-2}
			Action cost / constraint                      &  $c_a=1$, $c_b=0.8$, $c_c=0.5$, $\overline{c}_{\max}=0.23$     &  &  &    \\ 	\cline{1-2}
			Violation threshold              & $\zeta_0=0.1$ Hz, $\zeta_1=0.01$ Hz    &  &  &  \\ 	\cline{1-2}
			Conservative ratio for CRC             & $\theta=0.5$  &  &  &  \\ 	\cline{1-2}
		\end{tabular}
	\end{table}

	\begin{table*}[h]
		\centering
		\footnotesize
		\caption{Performance Comparison}
		\begin{tabular}{cc|c|c|c|c}
			\toprule
			\rowcolor[HTML]{FFFFFF} 
			\multicolumn{2}{c|}{\cellcolor[HTML]{FFFFFF}{\color[HTML]{000000} Algorithms}}                                 & {\color[HTML]{000000} \makecell{Violation probability}} & {\color[HTML]{000000} {\makecell{Normalized total cost }}} & {\color[HTML]{000000} \makecell{Updating cost}} & {\color[HTML]{000000} \makecell{Control cost}} \\ \midrule
			%		 			\rowcolor[HTML]{FFFFFF} 
			%		 			\multicolumn{1}{c|}{\cellcolor[HTML]{FFFFFF}}                                                  & Without any operation & 99.40\%                                       & 0.9287                              & 0.600                               & 0.3287                                  \\ \cline{2-6} 
			\rowcolor[HTML]{FFFFFF} 
			\multicolumn{1}{c|}{\cellcolor[HTML]{FFFFFF}}                                                  & RS\cite{9162973} & 21.49\%                                       & 0.627                               & 0.600                               & 0.027                                  \\ \cline{2-6} 
			\rowcolor[HTML]{FFFFFF} 
			\multicolumn{1}{c|}{\multirow{-1}{*}{\cellcolor[HTML]{FFFFFF}{\makecell{\textbf{Benchmarks}}}}}                                 & RR\cite{8995639}  & 15.25\%                                       & 0.629                               & 0.600                             & 0.029                                \\ \cline{2-6} 
			\rowcolor[HTML]{FFFFFF} 
			\multicolumn{1}{c|}{\cellcolor[HTML]{FFFFFF}}                                                &  AoI-aware\cite{8845114}  & 19.75\%                                       & 0.471                             & 0.444                       & 0.027                             \\ \cline{2-6} 
			\rowcolor[HTML]{FFFFFF} 
			\multicolumn{1}{c|}{\cellcolor[HTML]{FFFFFF}}                                                  &  AoII-aware\cite{9137714}  & 20.67\%                                       & 0.408                             & 0.382                       & 0.026                             \\ \cline{2-6} 
			\rowcolor[HTML]{E9EDF4} 
			\multicolumn{1}{c|}{\cellcolor[HTML]{FFFFFF}}                                                  & GSC  & 11.27\%                                       & 0.198                               & 0.167                                & 0.031                             \\ \cline{2-6} 
			\rowcolor[HTML]{E9EDF4} 
			\multicolumn{1}{c|}{\cellcolor[HTML]{FFFFFF}}                                                  & GSC+CAE  & 6.68\%                                       & 0.209                            & 0.167                             & 0.042                                \\ \cline{2-6} 
			\rowcolor[HTML]{E9EDF4} 
			\multicolumn{1}{c|}{\cellcolor[HTML]{FFFFFF}}                                                  & GSC+CAE+PC  & 1.74\%                                       & 0.216                             & 0.167                            & 0.049                               \\ \cline{2-6} 
			
			\rowcolor[HTML]{CBCEFB} 
			\multicolumn{1}{c|}{\multirow{-4}{*}{\cellcolor[HTML]{FFFFFF}{\makecell{\textbf{Proposed}\\\textbf{methods}}}}}   & GSC+CAE+CRC           & 0.71\%                                       & 0.221                             & 0.167                           & 0.054                                \\ 	\bottomrule
		\end{tabular}
	\end{table*}
	\begin{figure}[htbp]
		\centering
		{\includegraphics[height=3.45cm]{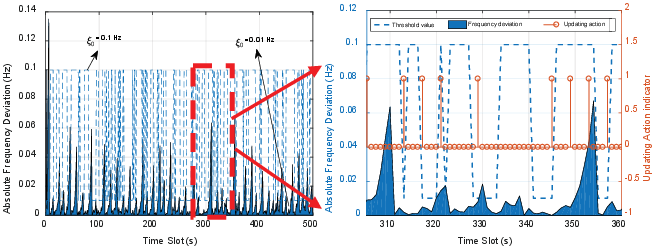}}
		\caption{Absolute value of frequency deviation with context-based threshold.}\label{fig_4}
	\end{figure}
	
	\subsection{Goal-oriented Co-Design Strategy}
	
	%Also, the transmission error probability is assumed to be the same for both uplinks.
	
	Based on the parameters provided in Table II, the violation probability and normalized cost obtained by different algorithms are shown in Table III.
	We assume that the transmitter
	has knowledge of the channel and source statistics.  
	Generally,  the proposed goal-oriented co-design policies can significantly reduce the violation probability and normalized cost simultaneously, thanks to the consideration of goal and context information. As shown in Fig. \ref{fig_4}, our proposed method can reduce the cost by remaining idle when the updating action is not necessary, and avoid violations by jointly considering goal-related state value and context information.
	%	    However, the existing work ignore the goal and its relation to actions, resulting in low transmission efficiency.
	Specifically,  by adopting the CAE scheme, the violation probability is reduced (up to 40\%) significantly with a slight increase in control cost (up to 5.5\%). 
	The estimation accuracy is improved and hence the control action is more efficient.
	Also, the sensing-aware control policies can  further reduce the violation probability by slightly increasing the number of control actions.
	Compared to the benchmarks, the proposed {GSC+CAE+CRC (use CRC hereafter)} policy can reduce the violation probability by up to 96.7\%, and reduce the total cost by up to 64.8\%.

	\subsection{Scenarios with Various Noise Variances and Downlink Error Probabilities}
	
\begin{figure}[htbp]
		\centering
		{\includegraphics[height=5cm]{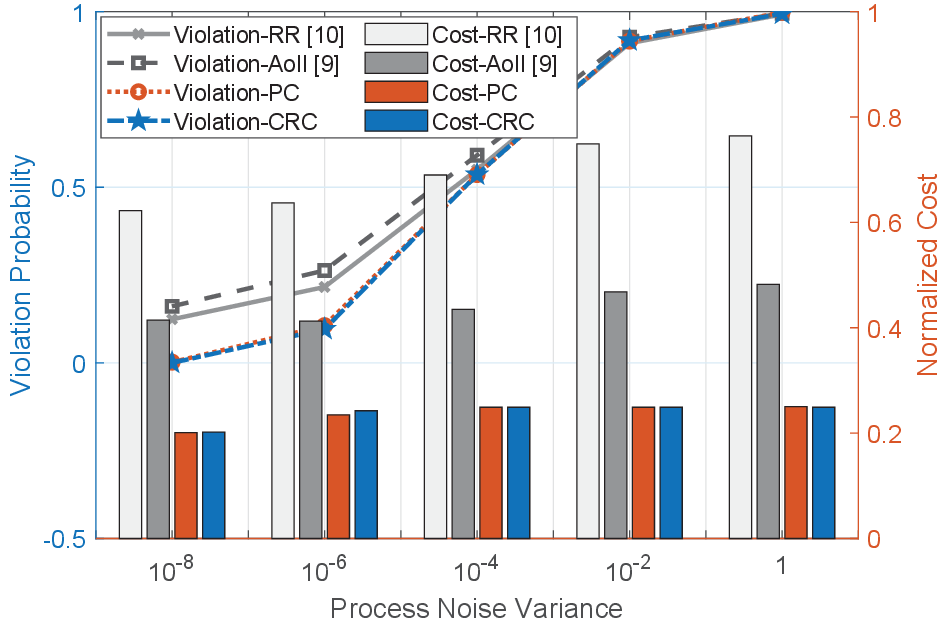}}
		\caption{Impact of process noise variance on the optimized  violation probability and and  cost.}\label{fig_5}
	\end{figure}

\begin{figure}[htbp]
		\centering
		{\includegraphics[height=5cm]{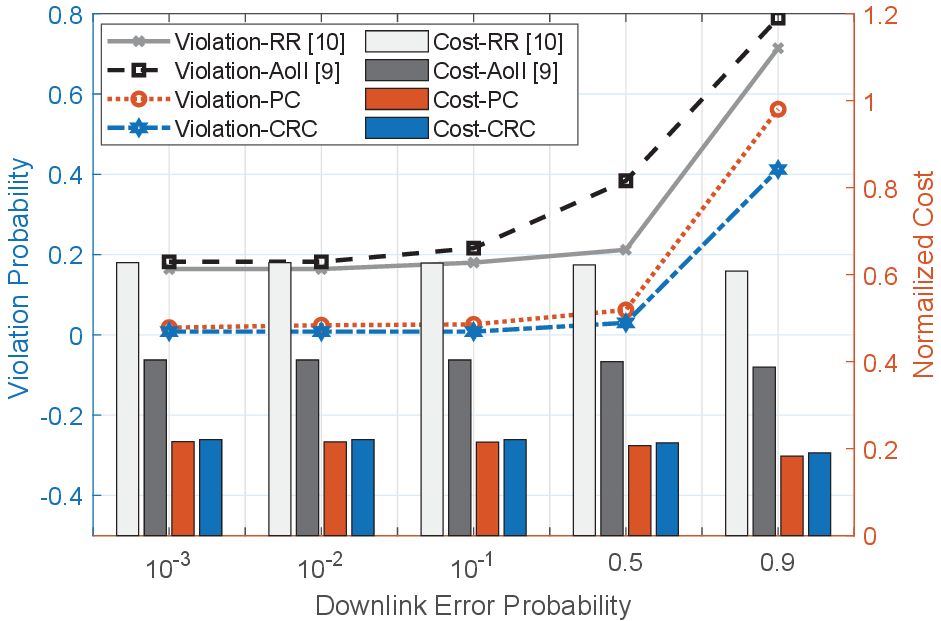}}
		\caption{Impact of downlink error probability on the optimized   violation probability and  cost.}\label{fig_6}
	\end{figure}

	We investigate the impacts of noise variance and error probability  on the violation probability and 
	total cost by different methods.
	As shown in Figs. \ref{fig_5} and \ref{fig_6}, the proposed policies can achieve the lowest violation probability (up to 44\% reduction compared to RR) and consume the lowest total cost (up to 50\% reduction compared to RR), especially with small noise variance and high error probability. 
	With a small noise variance,  accurate state estimation is obtained by the proposed PC and CRC methods,  leading to efficient control and low violation probability.
	%  	Also, Fig. 6 demonstrates that the proposed methods are robust to the error probability. 
	Moreover, the	CRC method
	achieves a lower violation probability with a trivial increase in cost, as the PC method  relies on the state estimation, which may not be accurate and results in  unstable control.

	
	\begin{figure}[htbp]
		\centering
		{\includegraphics[height=5.1cm]{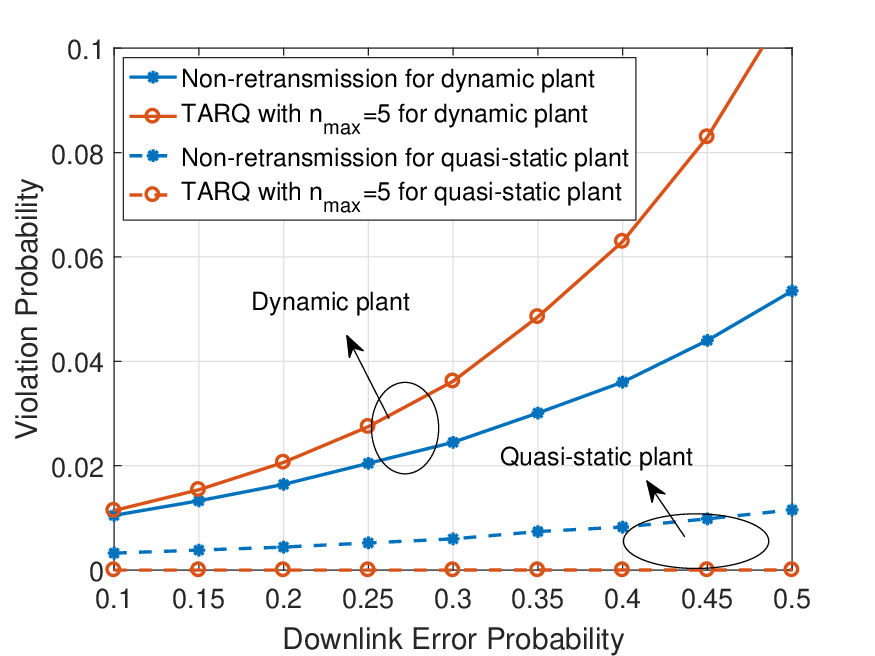}}
		\caption{Comparison between retransmission and non-retransmission schemes with different error probabilities and system dynamics.}\label{fig_7}
	\end{figure}

	\begin{figure*}[htbp]
		\centering
		{\includegraphics[height=4.2cm]{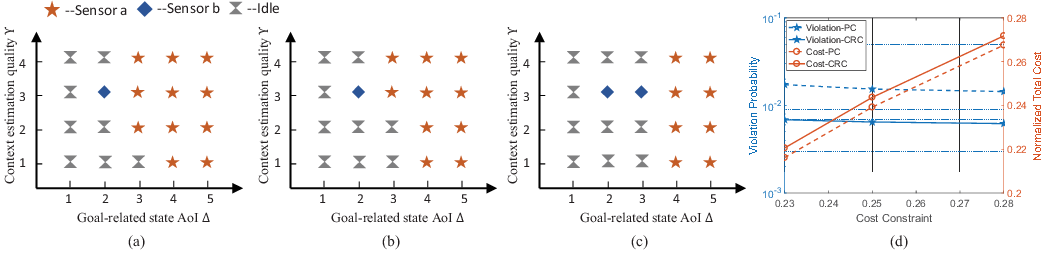}}
		\caption{Impact of cost constraint  on the scheduling policy with (a) $\overline{c}_{\max}=0.28$, $\lambda=0.5$,  (b) $\overline{c}_{\max}=0.25$, $\lambda=0.6$, (c) $\overline{c}_{\max}=0.23$, $\lambda=0.9$.}\label{fig_8}
	\end{figure*}
	\begin{figure*}[htbp]
		\centering
		{\includegraphics[height=4.15cm]{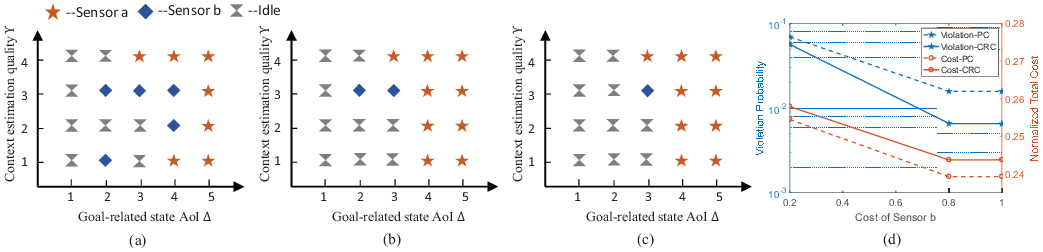}}
		\caption{Impact of sensor $S_b$'s cost  on the scheduling policy with (a) $c_b=0.2$, (b) $c_b=0.8$, (c) $c_b=1$.}\label{fig_9}
	\end{figure*}

	Fig. \ref{fig_7} investigates the impact of downlink error probability, where the CRC scheme is compared with the non-retransmission scheme.
	It is shown that retransmission is not always beneficial for reducing violation probability. For the  dynamic plant, the TARQ scheme achieves a higher violation probability, compared to the non-retransmission scheme. In the contrast, for the  quasi-static plant (system matrix is set to $\mathbf{A}=[0.08\quad1.1\quad0; 0\quad0\quad0.25; 0\quad0\quad1.1]$), the retransmission scheme can obtain a lower violation probability, as stated in Proposition 2.
	Furthermore, the gap between these two schemes is increasing with respect to the error probability.
	This is because the higher the error probability, the more frequently the TARQ scheme is triggered, leading to worse and better performance in dynamic and quasi-static scenarios, respectively.

	\subsection{Scenarios with Different Cost Constraints, Action Costs and  Error Probabilities for Uplink Transmission}
	Figs. \ref{fig_8} and \ref{fig_9} show the optimized scheduling policy, as well as the associated violation probabilities and costs with different cost constraints and action costs. 
	It is concluded that with a larger cost constraint, the Lagrangian multiplier is smaller, which leads to a more active scheduling policy and a lower violation probability, as analyzed in Subsection IV-C. Similarly, a lower $c_b$ increases the potential of transmitting sensor $S_b$ in the proposed scheduling policy.
	Interestingly, both violation probability and total cost decrease with respect to the sensor $S_b$'s cost. This is due to that scheduling sensor $S_b$ too frequently  may lead to insufficient goal-related state information and lack of control. 
	
\begin{figure}[htbp]
		\centering
		{\includegraphics[height=5.2cm]{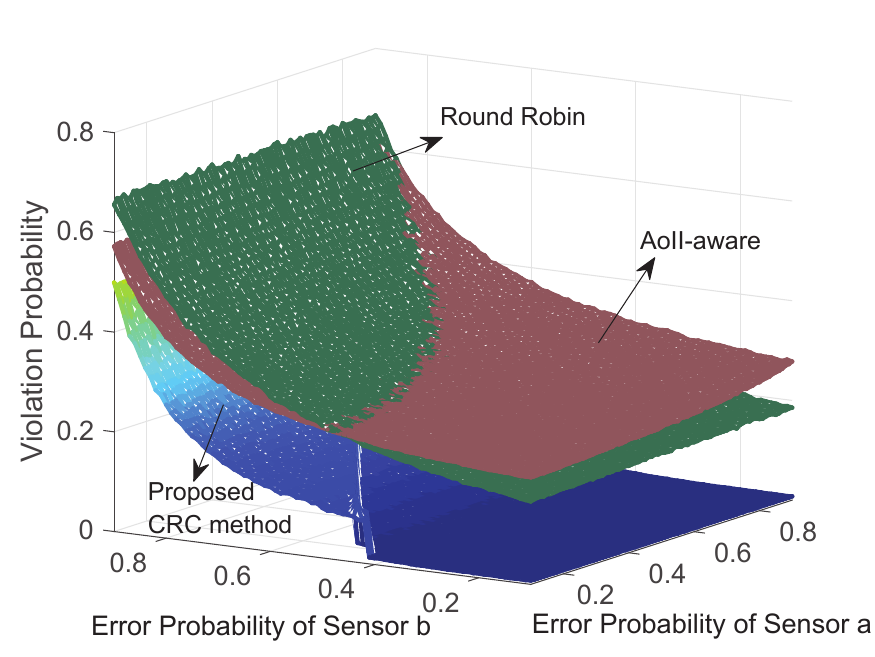}}
		\caption{Impact of uplink error probability on the optimized violation probability.}\label{fig_10}
	\end{figure}

	Fig. \ref{fig_10}  presents the impact of uplink error probability on the violation probability. It is demonstrated that the proposed co-design scheme achieves the best performance compared to the existing work. Also, the proposed methods are robust to the transmission error.	With the proposed CRC method, the violation probability remains low even with a relatively high error probability (\emph{e.g.}, $\epsilon_a=0.6$, $\epsilon_b=0.4$), thanks to accurate remote estimation and robust control.

	\begin{figure}[htbp]
		\centering
		{\includegraphics[height=5.2cm]{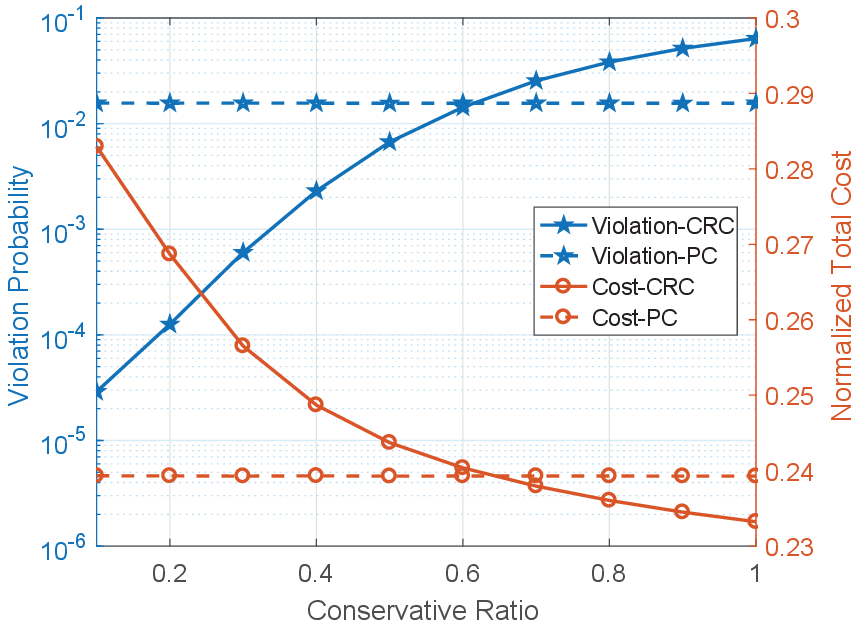}}
		\caption{Impact of conservative ratio on the optimized violation probability and cost.}\label{fig_11}
	\end{figure}

	Fig. \ref{fig_11} compares the proposed two methods with different conservative ratios.
	It is shown that the violation probability obtained by the CRC-based method is monotonically increasing with respect to the conservative ratio, while the total cost shows  the opposite trend.
	This is due to that with a smaller conservative ratio, the CRC-based method is more sensitive to the goal-related plant state, leading to more frequent control.
	Fig. \ref{fig_11} also implies the tradeoff between violation probability and cost. Appropriate conservative ratios can be selected according to different situations and requirements.
	
	\section{Conclusion}				
	In this paper, we have investigated the goal-oriented estimation, scheduling, and control for the WNCS with heterogeneous sources and two imperfect links. In particular, both goal-related state value and context information are monitored by multiple sensors and transmitted to the event-triggered controller for calculating control commands. 
	We have provided a comprehensive analysis of the goal of WNCS with respect to the scheduling and control actions and revealed their relationships. Then,  the co-design of scheduling and control has been formulated as a CMDP and simplified as an MDP using the Lagrangian relaxation method. Based on the structural results of MDP, we have proposed a goal-oriented deterministic scheduling policy to minimize the violation probability with the cost constraints.
	Furthermore, control-aware estimation and sensing-assisted control policies have been proposed to reduce the violation probability further. To further improve the transmission reliability and enhance the control stability, we have compared the retransmission scheme with the non-retransmission case and analyzed their preference scenarios.   Simulation results have shown that our proposed methods can achieve up to $96.7\%$ violation probability reduction and up to $64.8\%$  cost reduction compared to the existing scheduling methods. Moreover, the impacts of system parameters on the scheduling policy and the optimized violation probability and cost have been investigated comprehensively. Simulation results have demonstrated that the proposed methods are robust to the transmission error.
	
	In the future, we will consider a more practical WNCS with two-way random delay and imperfect feedback links. Furthermore,  more heterogeneous traffic sources and advanced control strategies will be included in future extensions of this work.

	\appendices
	
	\section{State Transition Probabilities}				
	With 	$s=(\Delta,\Upsilon=2)$,  the transition probability is given by
	\begin{footnotesize}
		\begin{equation}
			P(s'|s,\alpha)=\begin{cases}
				\epsilon_a\bar{p}, & if\quad \alpha=1, s'=(\Delta+1,\Upsilon=2) \\
				\epsilon_ap, & if\quad \alpha=1, s'=(\Delta+1,\Upsilon=4)  \\
				(1-\epsilon_a)\bar{p}, & if\quad \alpha=1, s'=(1,\Upsilon=2) \\
				(1-\epsilon_a)p, & if\quad \alpha=1, s'=(1,\Upsilon=4) \\
				(1-\epsilon_b)\bar{p}, & if\quad \alpha=2, s'=(\Delta+1,\Upsilon=1) \\
				(1-\epsilon_b)p, & if\quad \alpha=2, s'=(\Delta+1,\Upsilon=3) \\
				\epsilon_b\bar{p}, & if\quad \alpha=2, s'=(\Delta+1,\Upsilon=2) \\
				\epsilon_bp, & if\quad \alpha=2, s'=(\Delta+1,\Upsilon=4) \\
				\bar{p}, & if\quad \alpha=3, s'=(\Delta+1,\Upsilon=2 ) \\
				{p}, & if\quad \alpha=3, s'=(\Delta+1,\Upsilon=4) \\	 
				0, &otherwise.
			\end{cases}\label{eq_29}
		\end{equation}
	\end{footnotesize}
	With 	$s=(\Delta,\Upsilon=3)$,  the transition probability is given by
	\begin{footnotesize}
		\begin{equation}
			P(s'|s,\alpha)=\begin{cases}
				\epsilon_a\bar{p}, & if\quad \alpha=1, s'=(\Delta+1,\Upsilon=3) \\
				\epsilon_ap, & if\quad \alpha=1, s'=(\Delta+1,\Upsilon=1)  \\
				(1-\epsilon_a)\bar{p}, & if\quad \alpha=1, s'=(1,\Upsilon=3) \\
				(1-\epsilon_a)p, & if\quad \alpha=1, s'=(1,\Upsilon=1) \\
				(1-\epsilon_b)\bar{p}, & if\quad \alpha=2, s'=(\Delta+1,\Upsilon=4) \\
				(1-\epsilon_b)p, & if\quad \alpha=2, s'=(\Delta+1,\Upsilon=2) \\
				\epsilon_b\bar{p}, & if\quad \alpha=2, s'=(\Delta+1,\Upsilon=3) \\
				\epsilon_bp, & if\quad \alpha=2, s'=(\Delta+1,\Upsilon=1) \\
				\bar{p}, & if\quad \alpha=3, s'=(\Delta+1,\Upsilon=3 ) \\
				{p}, & if\quad \alpha=3, s'=(\Delta+1,\Upsilon=1) \\	 
				0, &otherwise.
			\end{cases}\label{eq_30}
		\end{equation}	
	\end{footnotesize}
	With 	$s=(\Delta,\Upsilon=4)$,  the transition probability is given by
	\begin{footnotesize}
		\begin{equation}
			P(s'|s,\alpha)=\begin{cases}
				\epsilon_a\bar{p}, & if\quad \alpha=1, s'=(\Delta+1,\Upsilon=4) \\
				\epsilon_ap, & if\quad \alpha=1, s'=(\Delta+1,\Upsilon=2)  \\
				(1-\epsilon_a)\bar{p}, & if\quad \alpha=1, s'=(1,\Upsilon=4) \\
				(1-\epsilon_a)p, & if\quad \alpha=1, s'=(1,\Upsilon=2) \\
				(1-\epsilon_b)\bar{p}, & if\quad \alpha=2, s'=(\Delta+1,\Upsilon=4) \\
				(1-\epsilon_b)p, & if\quad \alpha=2, s'=(\Delta+1,\Upsilon=2) \\
				\epsilon_b\bar{p}, & if\quad \alpha=2, s'=(\Delta+1,\Upsilon=4) \\
				\epsilon_bp, & if\quad \alpha=2, s'=(\Delta+1,\Upsilon=2) \\
				\bar{p}, & if\quad \alpha=3, s'=(\Delta+1,\Upsilon=4 ) \\
				{p}, & if\quad \alpha=3, s'=(\Delta+1,\Upsilon=2) \\	 
				0, &otherwise.
			\end{cases}\label{eq_31}
		\end{equation}		
	\end{footnotesize}

	\section{Proof of Proposition 2}				
	Suppose that sensor $S_a$ is scheduled at time slot $k$. Then we have $x_k$ and $\hat{x}_{k+1}$.	After $n-1$  retransmissions, the plant state is given by	
	$\mathbf{x}^{\text{ARQ}}_{k+1+n}
	%=&\mathbf{A}\mathbf{x}_{k+n}-\mathbf{A}\hat{\mathbf{x}}_{k+n}\\&+ \mathbf{A}\hat{\mathbf{x}}_{k+n}-\mathbf{A}\hat{\mathbf{x}}_{k+1}+\mathbf{w}_{k+n}\\
	=\mathbf{A}\mathbf{e}_{k+n}+\mathbf{w}_{k+n}+(\mathbf{A}^{n+1}-\mathbf{A}^{2}){\mathbf{x}}_{k}$.
	The plant-state covariance  is given by
	\begin{equation}
		\mathbf{\Phi}_{k+1+n}^{\text{ARQ}}=\begin{cases}
			\mathbf{A}\mathbf{\Theta}_{k+n}\mathbf{A}^{\top}+\mathbf{R_w}+	\mathbf{\Sigma}_{k,n+1}, & \omega_{c,k+n}=1, \\
			\mathbf{A}\mathbf{\Phi}_{k+n}\mathbf{A}^{\top}+\mathbf{R_w}, & otherwise,
		\end{cases}\label{eq_32}
	\end{equation}
	where $\mathbf{\Sigma}_{k,n+1}=(\mathbf{A}^{n+1}-\mathbf{A}^2)\mathbf{\Phi}_{k}(\mathbf{A}^{n+1}-\mathbf{A}^2)^{\top}$.
	The expectation is given by
	\begin{equation}
		\begin{aligned}
			\mathbb{E}\left[\mathbf{\Phi}_{k+1+n}\right]^{\text{ARQ}}=&(1-\epsilon^{n})Z_1+\epsilon^{n}Z_2.
		\end{aligned}\label{33}
	\end{equation}
	where $Z_1=\mathbf{A}\mathbf{\Theta}_{k+n}\mathbf{A}^{\top}+\mathbf{R_w}+\mathbf{\Sigma}_{k,n+1}$, and $Z_2=\mathbf{A}\mathbf{\Phi}_{k+n}\mathbf{A}^{\top}+\mathbf{R_w}$.		
	
	In comparison, 
	with $n-1$ transmission under non-retransmission policy, we have 
	$\mathbf{x}_{k+1+n}^{\text{Non}}=\mathbf{A}\mathbf{e}_{k+n}+\mathbf{w}_{k+n}$.
	Then the expectation of state covariance is given by
	\begin{equation}
		\begin{aligned}
			\mathbb{E}\left[\mathbf{\Phi}_{k+1+n}\right]^{\text{Non}}=&(1-\epsilon)(Z_1-\mathbf{\Sigma}_{k,n+1})+\epsilon Z_2.
		\end{aligned}\label{eq_34}
	\end{equation}
	Based on (\ref{eq_32}) and (\ref{eq_34}), the gap between two schemes can be obtained, as  in (\ref{eq_27}). 
	
	Then we analyze the relationship between the gap and system matrix as well as the error probability. First, for the quasi-static plant, \emph{i.e.}, $\rho(\mathbf{A})\rightarrow1$, we have $\mathbf{\Sigma}_{k,n+1}\rightarrow0$. 		In this case, the second term in (\ref{eq_27}) approaches 0. Hence, $\mathbb{E}\left[D\right]<0$ holds since $(\mathbf{A}\mathbf{\Theta}_{k+n}\mathbf{A}^{\top}-\mathbf{A}\mathbf{\Phi}_{k+n}\mathbf{A}^{\top})=-\mathbf{A}^{n+1}\mathbf{\Phi}_{k}{\mathbf{A}^{n+1}}^{\top}<0$ is true based on  (\ref{eq_6}) and (\ref{eq_11}).
	Furthermore, with a larger error probability, $(\epsilon-\epsilon^n)$ is larger, leading to a smaller $\mathbb{E}\left[D\right]$.
	For the dynamic plant, \emph{i.e.}, $\rho(\mathbf{A})\gg1$, we have 
	$\mathbf{\Sigma}_{k,n+1}\approx \mathbf{A}^{n+1}\mathbf{\Phi}_{k}{\mathbf{A}^{n+1}}^{\top}$. Therefore, we have  $\mathbb{E}\left[D\right]>0$ based on the fact that $\mathbf{\Sigma}_{k,n+1}>0$ and $\epsilon<1$.
	Moreover, in this case, a larger gap is obtained with a larger error probability. This is due to that a larger error probability also leads to a larger number of retransmissions, resulting in a larger $\mathbf{\Sigma}_{k,n+1}$ and hence a larger gap.

	\vfill
	
\end{document}